\documentclass[12pt]{article}
\usepackage{theorem}
\usepackage{amsfonts}
\newfont\fiverm{cmr5}
\usepackage{url}

\advance\textheight2.7cm        
\advance\topmargin-2.0cm
\advance\textwidth2.6cm
\advance\evensidemargin-1.6cm
\advance\oddsidemargin-1.6cm

\theoremstyle{change}
\newtheorem{thm}{Theorem.\nopagebreak}[section]

\newtheorem{cor}[thm]{Corollary.\nopagebreak}

\newtheorem{prop}[thm]{Proposition.\nopagebreak}

\theorembodyfont{\rmfamily}
\newtheorem{dfn}[thm]{Definition.\nopagebreak}

\newtheorem{expl}[thm]{Example.\nopagebreak}

\newtheorem{expls}[thm]{Examples.\nopagebreak}

\newtheorem{rems}[thm]{Remarks.\nopagebreak}

\newenvironment{proof}{{\it Proof. }}{~~~\hfill$\Box$\vspace{0.5cm}}
\def\bepf{\begin{proof}}
\def\epf{\end{proof}}

\def\fct#1{\mathop{\rm #1}}	
\def\fns#1{{\mbox{\rm \scriptsize#1}}} 		

\def\D{\displaystyle}				
\def\ol{\overline}

\def\hbar{{\mathchar'26\mkern-9muh}}
\def\kbar{{\mathchar'26\mkern-9muk}}
\def\sint{\mbox{\small $\int$}}

\def\downto{\downarrow}
\def\<{\langle} 				
\def\>{\rangle} 				
\def\lbeq#1{\begin{equation} \label{#1}} 
\def\eeq{\end{equation}} 
\def\bary{\begin{array}}
\def\eary{\end{array}}
\def\gzit#1{{\rm (\ref{#1})}} 			
\def\spd{\hspace*{1.0cm}} 			
\def\eps{\varepsilon}

\def\Lin{\fct{Lin}}
\def\tr{\fct{tr}}
\def\const{\fct{const}}
\def\kin{\fns{kin}}
\def\Forall{~~~\mbox{for all }}  
\def\implies{~~~\Rightarrow~~~}
\def\for{~~~\mbox{for }}
\def\half{\frac{1}{2}}

\def\Cz{\mathbb{C}}
\def\Rz{\mathbb{R}}
\def\Ez{\mathbb{E}}
\def\Hz{\mathbb{H}}

\def\res{\fct{Res}}

\makeindex

\begin{document}

\vspace*{-2cm}

\begin{center}

{\LARGE \bf On the foundations of thermodynamics}

\vspace{1cm}

\centerline{\sl {\large \bf Arnold Neumaier}}

\vspace{0.5cm}

\centerline{\sl Fakult\"at f\"ur Mathematik, Universit\"at Wien}
\centerline{\sl Nordbergstr. 15, A-1090 Wien, Austria}
\centerline{\sl email: Arnold.Neumaier@univie.ac.at}
\centerline{\sl WWW: \url{http://www.mat.univie.ac.at/~neum}}

May 25, 2007
\end{center}


\vspace{0.5cm}

{\bf Abstract.} 
On the basis of new, concise foundations, this paper establishes 
the four laws of thermodynamics, the Maxwell relations, and the 
stability requirements for response functions, in a form applicable to 
global (homogeneous), local (hydrodynamic) and microlocal (kinetic) 
equilibrium.

The present, self-contained treatment needs very little formal 
machinery and stays very close to the formulas as they are applied 
by the practicing physicist, chemist, or engineer. 
From a few basic assumptions, the full structure of phenomenological 
thermodynamics and of classical and quantum 
statistical mechanics is recovered.

Care has been taken to keep the foundations free of subjective aspects 
(which traditionally creep in through information or probability). 
One might describe the paper as a uniform treatment of the nondynamical 
part of classical and quantum statistical mechanics 
``without statistics'' (i.e., suitable for the definite descriptions 
of single objects) and 
``without mechanics'' (i.e., independent of microscopic assumptions).
When enriched by the traditional examples and applications, this paper 
may serve as the basis for a course on thermal physics.

\vfill
\begin{flushleft}
{\bf Keywords}: 
density, 
energy balance,
energy dissipation,  
entropy, 
equation of state,
equilibrium state, 
equilibrium thermodynamics, 
Euclidean *-algebra,
Euler equation, 
exponential family,
extremal principles,
foundations of thermodynamics,
free energy,
Gibbs paradox,
Gibbs state, 
Gibbs-Bogoliubov inequality,
information deficit,
Kubo inner product,
Legendre transform,
limit resolution,
partition function,
quantization,
response functions,
Schr\"odinger equation,
second law of thermodynamics, 
thermal interpretation of quantum mechanics,
thermal state,
thermodynamic stability,
unobservable complexity
\end{flushleft}


{\bf E-print Archive No.}: cond-mat/yymmnnn\\
{\bf 2006\hspace{.4em} PACS Classification}: primary 05.70.-a

\newpage
\tableofcontents 

\newpage 
\section{Introduction} \label{intro}

It is worthwhile to have the structural properties of physical 
theories spelled out independent of computational techniques for 
obtaining quantitatively correct numbers from the assumptions made. 
This allows one to focus attention on the simplicity and beauty of 
theoretical physics, which is often hidden in a jungle of techniques 
for estimating or calculating approximations for quantities of 
interests. 
The standard approximation machinery for calculating explicit 
thermodynamic properties of materials from first principles can be 
postponed to a later stage where one actually wants to do quantitative 
predictions of macroscopic properties from microscopic assumptions. 

In the present paper, a treatment of equilibrium statistical mechanics
and the kinematic part of nonequilibrium statistical mechanics
is given which shows that such a structural point of view leads to 
great conceptual simplifications:
From a single basic assumption (Definition \ref{3.1.}), the full 
structure of phenomenological thermodynamics and 
of statistical mechanics is recovered (except for the third law 
which requires an additional quantization assumption), compatible with
both a classical or a quantum microscopic view, but without the 
elaborate formalism needed to do actual statistical mechanics 
calculations. (In this paper, these are only hinted at, as a comment 
on the cumulant expansion.)

Our treatment is self-contained and stays very close to the formulas 
as they are applied by the practicing physicist, chemist, or engineer.
It needs very little formal machinery, in contrast to other 
axiomatic settings for thermodynamics (e.g., 
{\sc Lieb \& Yngvason} \cite{LieY}). A different, well-known axiomatic
foundation of (classical) thermodynamics was given by 
{\sc Carath\'eodory} \cite{Car}. See also {\sc Emch \& Liu} \cite{EmcL}.

In contrast to traditional treatments of thermodynamics 
(e.g., {\sc Haken} \cite{Haken}, {\sc Jaynes} \cite{Jaynes}, 
{\sc Katz} \cite{Kat}), we do not {\it assume} 
the basic thermodynamical laws and relations but {\it derive}
them from simple definitions. In our treatment, the zeroth law is
made precise as a definition, and each of the other three laws 
becomes a theorem. 
All formulas are valid rigorously without approximation, assuming
only that the states in question are thermal states.

Since thermodynamics is completely observer independent,
its foundations should have this feature, too.
Therefore, care has been taken to keep the foundations free of 
subjective aspects, traditionally introduced by an information 
theoretic approach (see {\sc Balian} \cite{Bal3} for a recent 
exposition) to statistical mechanics in terms of the knowledge of an 
observer.
The present approach shows that (and in which sense) a fully objective, 
observer-independent foundation is possible. 

One might describe the paper as a uniform treatment of the nondynamical 
part of classical and quantum statistical mechanics 
``without statistics and probability'' (i.e., suitable for the definite 
descriptions of single objects) and 
``without mechanics'' (i.e., independent of microscopic assumptions).
Essenitally, the paper develops all relevant results without assuming 
or claiming anything definite about the meaning of the quantities 
assumed to exist on the fundamental description level. Thus the 
interpretation of an underlying microscopic structure is left 
completely open; even the 'number of particles' can be taken 
figuratively!

Avoiding a definite mechanical picture frees the foundations from the 
unresolved (and in principle unresolvable) issues regarding the truly 
fundamental constituents of nature and possibly needed modifications of 
quantum field theory at the Planck scale where gravitational effects 
must be taken into account. Avoiding probability implies that the 
traditional foundational problems of quantum mechanics 
-- the most informative sources include 
{\sc Stapp} \cite{Sta}, {\sc Ballentine} \cite{Ball},
{\sc Home \& Whitaker} \cite{HomW}, {\sc Peres \& Terno} \cite{PerT} 
and the reprint collection by {\sc Wheeler \& Zurek} \cite{WheZ} --
and of equilibrium statistical mechanics, 
carefully analyzed by {\sc Sklar} \cite{Skl} for classical 
mechanics, (see also {\sc Grandy} \cite{Gran} and the older surveys
by {\sc Ehrenfest} \cite{Ehr}, 
{\sc ter Haar} \cite{tHaa}, {\sc Penrose} \cite{Pen}), and by
{\sc Wallace} \cite{Wal} for quantum mechanics,
do not affect the foundations of thermodynamics.

\bigskip
This is one of a sequence of papers designed to give a mathematically 
precise and philosophically consistent foundation of
modern theoretical physics. The ultimate goal are foundations of 
physics which are fully objective, mathematically concise, and
describe the universe as a whole, including at times when human 
observers could not exist. The first paper in the series 
({\sc Neumaier} \cite{Neu.ens}) defines and discusses the concepts of 
quantities, ensembles, and experiments in classical and quantum physics.

On this basis (but formally independent), the present paper derives the 
standard formalism of equilibrium thermodynamics and statistical 
mechanics, formulated in a way that includes global equilibrium 
(homogeneous materials), 
local equilibrium (hydrodynamics and continuum mechanics), 
microlocal equilibrium (kinetic descriptions), and quantum equilibrium 
(microscopic descriptions). In this paper, only the kinematical level
is discussed; another paper ({\sc Neumaier} \cite{Neu.thd}) on the
dynamics of thermal states is in preparation.
The formalism developed is closely related to the theory of exponential 
families (cf. {\sc Barndorff-Nielsen} \cite{BarN,BarN2}, 
{\sc Bernardo \& Smith} \cite{BerS}), although the language used there 
is completely different.

The physics of our results is, of course, well-known since 
{\sc Gibbs} \cite{Gib}; see {\sc Uffink} \cite{Uff} for a history. 
However, the treatment of the material is new,
as is the careful set-up used to arrive at the standard results. 
In particular: We make precise (in Section \ref{s.phen}) to which 
extent thermodynamical variables have a meaning outside equilibrium, 
thus ensure a formally well-defined second law.   
We give (in Section \ref{s.limit}) 
a new interpration of the notion of an ensemble, valid for single 
macroscopic systems, whose properties are well-defined 
without reference to multiple realizations, and clarify the role of 
statistics in the foundations. We give (in Sections
\ref{zeroth},  \ref{first},  \ref{second}, and  \ref{third})
concise formal statements of the fundamental laws of thermodynamics. 
Finally, we demonstrate (in Section \ref{third}) 
that the conventional wave function form of quantum mechanics can be 
simply understood as the low temperature limit of thermodynamics.

\bigskip
The paper is organized as follows:
To set the physical background needed to appreciate the remainder 
of the paper, Section \ref{s.phen} and Section \ref{s.cons}
give a description 
of standard phenomenological equilibrium thermodynamics for 
single-phase systems in the absence of chemical reactions and 
electromagnetic fields. The present, unconventional approach summarizes
the assumptions needed in a way that allows an easy comparison 
with statistical mechanics.
Section \ref{s.gibbs} introduces Gibbs states, in a generality 
covering both the classical and the quantum case; their abstrct 
properties are studied in Section \ref{s.gen}. 
Section \ref{s.limit} discusses the role of statistics in our
approach.
Section \ref{zeroth} defines thermal states and discusses their
relevance for global, local, and microlocal equilibrium. 
Section \ref{s.eos} deduces the existence of an equation of state
and connects the results to the phenomenological
exposition in Section \ref{s.phen}.
Section \ref{s.detail} discusses questions relating to different
thermal description levels.
Section \ref{first} proves the first law of thermodynamics.
In Sections \ref{second}, we compare 
thermal states with arbitrary Gibbs states and deduce the
extremal principles of the second law.
Section \ref{third} shows that the third law is related to 
a simple quantization condition for the entropy and relates it
to the time-independent Schr\"odinger equation.
Section \ref{s.model} discusses the hierarchy of equilibrium
descriptions and how they relate to each other.

Appendix \ref{s.complexity} exhibits the relations to information 
theory, by putting a simple decision problem into our framework. 
In this example, we recover the usual interpretation of the 
value of the entropy as a measure of unobservable internal complexity 
(lack of information).
Appendix \ref{s.maxent} discusses the maximum entropy principle and 
its limitations.
Appendix \ref{s.technical} contains the proofs of a few auxiliary
results needed in the main text. 

In this paper, we shall denote quantities with a specific meaning 
by capital letters, conforming to the tradition in thermodynamics.

The level of formal precision is somewhere between that of mathematical
physics and that of most treatments of theoretical physics. 
I made my best efforts to carefully argue all relevant issues and to
make explicit all assumptions used in these argument.
Definitions, theorems, and proofs are stated precisely.
However, to avoid an excessive mathematical apparatus 
we do not specify the domains of definition of functions, and
we state but do not use axioms relating to continuity properties and 
similar things, though these would be necessary for a completely 
rigorous treatment. If one assumes that the algebra of quantities 
is finite-dimensional, all proofs are valid rigorously
as stated. For the infinite dimensional case, these arguments need 
further justification. However, assuming the results in Appendix 
\ref{s.technical}, all other results still follow rigorously.
(In fact, it can be shown that everything would hold rigorously
when this algebra is embeddable into a C$^*$-algebra. 
But the applications require certain unbounded operators to occur.
Moreover, unlike our nearly trivial axioms, 
the C$^*$-algebra axioms have no intuitive physical basis, but
are abstracted from mathematical properties of algebras of bounded
operators on Hilbert spaces. Thus, we do not use the  C$^*$-algebra
framework.)

{\bf Acknowledgment.} Thanks to Roger Balian, Clemens Elster, 
Mike Mowbray, Hermann Schichl and Tapio Schneider 
for useful comments on earlier versions of this manuscript, and to 
Tom Epperly who, many years ago, has introduced me to thermodynamics.

\section{Phenomenological thermodynamics}\label{s.phen}

To introduce the physical background for our developments,
we discuss thermodynamic systems which describe the single-phase 
global equilibrium of matter in the absence of chemical reactions and 
electromagnetic fields. We call such systems {\bf standard
thermodynamic systems}; they are ubiquitous in the applications. 
In particular, a standard system is considered to be uncharged, 
homogeneous, and isotropic, so that each finite 
region looks like any other and is very large in microscopic units.
(Multiple phases would be only piecewise homogeneous; each phase 
separately is described as indicated, but discussing the equilibrium 
at interfaces needs some additional effort, described in all textbooks 
on thermodynamics.) 

Our phenomenological approach is similar to that of {\sc Callen} 
\cite{Cal}, a widely accepted authority on equilibrium thermodynamics, 
who introduces the basic concepts by means of a few postulates from 
which everything else follows. 
The present setting has the advantage that it
easily matches the more fundamental approach based on statistical 
mechanics. Our axioms are also slightly more precise than Callen's in 
that they specify the kinematical properties of states outside 
equilibrium. This enables us to replace his informal thermodynamic 
stability arguments (which depends on dynamical assumption close to 
equilibrium) by rigorous mathematical arguments.

The species of molecules of fixed chemical composition are labeled 
by an index $j\in J$; we speak of {\bf particles of species} $j$.
A standard thermodynamic system is characterized by the {\bf number 
$N_j$ of particles} of each species $j$, the corresponding 
{\bf chemical potential} $\mu_j$ of species $j$, the {\bf volume} $V$, 
the {\bf pressure} $P$, the {\bf temperature} $T$, the {\bf entropy} 
$S$, and the {\bf Hamilton energy} $H$. These variables, the 
{\bf extensive} variables $N_j,V,S,H$ and the {\bf intensive} variables 
$\mu_j,P,T$, are jointly called the {\bf basic thermodynamic variables}.
In the terminology, we mainly follow the IUPAC convention 
({\sc Alberty} \cite[Section 7]{Alb}), except that we use the letter
$H$ to denote the Hamilton energy, as customary in quantum mechanics. 
In equilibrium, $H$ equals the internal energy. The Hamilton energy 
should not be confused with the enthalpy which (is usually denoted by 
$H$ but here) is given in equilibrium by $H+PV$.
(For a history of thermodynamics notation, see 
{\sc Battino} et al. \cite{BatSW}.)

As customary in thermodynamics, we allow
the $N_j$ to have arbitrary nonnegative real values, not only integers.
(Chemists use instead of particle numbers the corresponding mole 
numbers, which differ by a fixed numerical factor, the Avogadro 
constant; cf. Example \ref{ex.ideal}.) 
We group the $N_j$ and the $\mu_j$ into vectors $N$ and $\mu$ indexed 
by $J$ and write $\mu\cdot N = \sum_{j\in J} \mu_jN_j$. 
If there is just a single species, we drop the indices and have 
$\mu\cdot n = \mu N$. In this section, all numbers are real.

Equilibrium thermodynamics is about characterizing so-called
equilibrium states and comparing them with similar nonequilibrium 
states. Everything of relevance in equilibrium thermodynamics can be 
deduced from the following

\begin{dfn}{\bf (Axioms for phenomenological thermodynamics)}
\label{d.phen}\\
(i) Temperature $T$, pressure $P$, volume $V$ are positive,
particle numbers $N_j$ are nonnegative. There is a convex\footnote{
See Appendix \ref{s.technical} for convexity and the associated 
differentiability almost everywhere. Surfaces of nondifferentiability
correspond to so-called {\bf phase transitions}. Although practically 
important and hence discussed in every textbook of thermodynamics, 
phase transitions do not raise foundational problems. Therefore,
we look in this paper only at regions of state space where $\Delta$ 
is twice continuously differentiable.
}\ 
{\bf system function} $\Delta$ of the intensive variables $(T,P,\mu)$ 
which is monotone increasing in $T$ and monotone decreasing in $P$, 
such that the intensive variables are related by the {\bf equation of
state}
\lbeq{e.def}
\Delta(T,P,\mu)=0.
\eeq
The set of $(T,P,\mu)$ satisfying $T>0$, $P>0$ and the equation of 
state is called the {\bf state space}. 

(ii) Outside equilibrium, only the extensive variables are well-defined;
they are additive under the composition of disjoint subsystems. 
The Hamilton energy $H$ satisfies the {\bf Euler inequality}
\lbeq{e.Hi}
H\ge T S - P V + \mu \cdot N
\eeq
for all $(T,P,\mu)$ in the state space.
{\bf Equilibrium states} have well-defined intensive and extensive 
variables, and are are defined by equality in \gzit{e.Hi}.
\end{dfn}

This is the complete list of assumptions which define phenomenological 
equilibrium thermodynamics; the system function $\Delta$ can be 
determined either by fitting to experimental data, or by calculation 
from more fundamental descriptions, cf. Theoren \ref{t.eos}.
All other properties follow from the system function.
Thus, the equilibrium properties of a material (e.g., ''salty water'',
$J=\{$salt,water$\}$) are characterized by the system function 
$\Delta$; each equilibrium instance of the material (''this glass 
of salty water'') is characterized by a particular state 
$(T,P,N)$, from which all equilibrium properties can be computed:

\begin{thm}~

(i) In any equilibrium state, the extensive variables are given by 
\lbeq{e.Sp}
S=\Omega\frac{\partial \Delta}{\partial T}(T,P,\mu),~~~
V=-\Omega\frac{\partial \Delta}{\partial P}(T,P,\mu),~~~
N=\Omega\frac{\partial \Delta}{\partial \mu}(T,P,\mu),
\eeq
and the {\bf Euler equation}
\lbeq{e.Hp}
H=T S - P V + \mu \cdot N,
\eeq
where $\Omega$ is a positive number called the {\bf system size}. 

(ii) In equilibrium, we have the {\bf Maxwell reciprocity relations}
\lbeq{7-10p}
-\frac {\partial V} {\partial T}
=\frac {\partial S} {\partial P},~~
\frac {\partial N_j} {\partial T}
=\frac {\partial S} {\partial\mu_j},~~
\frac {\partial N_j} {\partial P}
=-\frac {\partial V} {\partial\mu_j},~~
\frac {\partial N_j} {\partial\mu_k}
=\frac {\partial N_k} {\partial\mu_j}
\eeq
and the {\bf stability conditions}
\lbeq{7-13p}
\frac {\partial S} {\partial T}\ge 0,~~ 
\frac {\partial V} {\partial P}\le 0,~~ 
\frac {\partial N_j} {\partial\mu_j}\ge 0.
\eeq
\end{thm}

\bepf
In equilibrium, at fixed $S,V,N$, the triple $(T,P,\mu)$ is a maximizer
of $TS-PV+\mu\cdot N$ under the constraints $\Delta(T,P,\mu)=0$, $T>0$, 
$P>0$. A necessary condition for a maximizer is the stationarity of 
the Lagrangian
\[
L(T,P,\mu)=TS-PV+\mu\cdot N -\Omega\Delta(T,P,\mu)
\]
for some Lagrange multiplier $\Omega$. Setting the partial derivatives 
to zero gives \gzit{e.Sp}, and since the maximum is attained in 
equilibrium, the Euler equation \gzit{e.Hp} follows. The system size
$\Omega$ is positive since $V>0$ and $\Delta$ is decreasing in $P$.
Since the Hessian matrix of $\Delta$, 
\[
\Sigma= 
\left(\begin{array}{rrr}
\D\frac{\partial^2\Delta}{\partial T^2}& 
\D\frac{\partial^2\Delta}{\partial P\partial T}& 
\D\frac{\partial^2\Delta}{\partial \mu\partial T} \\
\D\frac{\partial^2\Delta}{\partial T\partial P}& 
\D\frac{\partial^2\Delta}{\partial P^2}& 
\D\frac{\partial^2\Delta}{\partial \mu\partial P} \\
\D\frac{\partial^2\Delta}{\partial T \partial \mu}& 
\D\frac{\partial^2\Delta}{\partial P \partial \mu}& 
\D\frac{\partial^2\Delta}{\partial\mu^2}
\end{array}\right)
= \Omega^{-1}
\left(\begin{array}{rrr}
\D\frac{\partial S}{\partial T}& 
\D\frac{\partial S}{\partial P}& 
\D\frac{\partial S}{\partial \mu} \\
\D-\frac{\partial V}{\partial T}& 
\D-\frac{\partial V}{\partial P}& 
\D-\frac{\partial V}{\partial \mu} \\
\D\frac{\partial N}{\partial T}& 
\D\frac{\partial N}{\partial P}& 
\D\frac{\partial N}{\partial\mu}
\end{array}\right),
\]
is symmetric, the Maxwell reciprocity relations follow, and since 
$\Delta$ is convex, $\Sigma$ is positive semidefinite; hence the 
diagonal elements of $\Sigma$ are nonnegative, giving the stability 
conditions.
\epf

Note that there are further stability conditions since the determinants
of all principal submatrices of $\Sigma$ must be nonnegative. Also, 
$N_j\ge 0$ implies that $\Delta$ is monotone increasing in each $\mu_j$.

\begin{expl}\label{ex.ideal}
The equilibrium behavior of electrically neutral gases at sufficiently 
low pressure can be modelled as ideal gases.
An {\bf ideal gas} is defined by a system function of the form
\lbeq{e.ideal}
\Delta(T,P,\mu)=\sum_{j\in J}\pi_j(T) e^{\mu_j/\kbar T}-P,
\eeq
where the $\pi_j(T)$ are positive functions of the temperature,
$\kbar$ is the {\bf Boltzmann constant} and we use the 
bracketing convention $\mu_j/\kbar T =\mu_j/(\kbar T)$.
The Boltzmann constant defines the units in which the entropy is 
measured. By a change of units one can enforce any value of $\kbar$.
In traditional macroscopic units, the Boltzmann constant is very tiny
\lbeq{e.kbar}
\kbar \approx 1.38065 \cdot 10^{-23} J/K.
\eeq
(For this and other constants, see {\sc CODATA} \cite{CODATA}.
In analogy with Planck's constant $\hbar$,
we write $\kbar$ in place of the customary $k$ or $k_B$, in order to 
be free to use the letter $k$ for other purposes.)
Differentiation with respect to $P$ shows that $\Omega=V$ is the 
system size, and from \gzit{e.def}, \gzit{e.Sp}, and \gzit{e.int}, 
we find that, in equilibrium,
\[
P=\sum_j \pi_j(T) e^{\mu_j/\kbar T},~~~
S=V\sum_j \Big(\frac{\partial}{\partial T}\pi_j(T) 
-\frac{\mu_j\pi_j(T)}{\kbar T^2}\Big) e^{\mu_j/\kbar T},
\]
\[
N_j =\frac{V\pi_j(T)}{\kbar T} e^{\mu_j/\kbar T},~~~
H=V\sum_j \Big(T\frac{\partial}{\partial T}\pi_j(T) 
-\pi_j(T)\Big) e^{\mu_j/\kbar T}.
\]
Expressed in terms of $T,V,N$, we have
\[
PV=\kbar T \sum_j N_j,~~~\mu_j=\kbar T\log\frac{\kbar T N_j}{V\pi_j(T)},
\]
\[
H=\sum_j h_j(T)N_j,~~~
h_j(T)=\kbar T\Big(T\frac{\partial}{\partial T}\log\pi_j(T) -1\Big),
\]
from which $S$ can be computed by means of the Euler equation 
\gzit{e.Hp}. 
In particular, for one {\bf mole} of a single substance, defined by 
$N=R/\kbar \approx 6.02214 \cdot 10^{23}$ (the 
{\bf Avogadro constant}), where $R\approx 8.31447\, J/K$ is the 
{\bf molar gas constant}, we get the {\bf ideal gas law} 
\lbeq{e.ilaw}
PV=RT;
\eeq
cf. {\sc Clapeyron} \cite{Clap}, {\sc Jensen} \cite{Jen}.

In general, the difference $h_j(T)-h_j(T_0)$ can be found 
experimentally by measuring the energy needed for raising or lowering 
the temperature of pure substance $j$ from $T_0$ to $T$ while keeping 
the $N_j$ constant. In terms of infinitesimal increments, the 
{\bf heat capacities} $C_V(T)=dh_j(T)/dT$,
we have
\[
h_j(T)=h_j(T_0)+\int_{T_0}^T dT\, C_V(T).
\]
From the definition of $h_j(T)$, we find that
\[
\pi_j(T)=\pi_j(T_0)\exp \int_{T_0}^T \frac{dT}{T}
\Big(1+\frac{h_j(T)}{\kbar T}\Big).
\]
Thus there are two undetermined integration constants for each 
particle species. These cannot be determined experimentally as long
as we are in the range of validity of the ideal gas approximation.
Indeed, if we pick arbitrary constants $p_j$ and $q_j$ and replace
$\pi_j(T),\mu_j,H$, and $S$ by 
\[
\pi_j'(T):=e^{p_j-q_j/\kbar T}\pi_j(T),~~~\mu_j'=\mu_j+q_j-\kbar T p_j,
\]
\[
H'=H+\sum_j p_jN_j,~~~S'=S+\kbar \sum_j q_jN_j,
\]
all relations remain unchanged. Thus, the Hamilton energy and the 
entropy of an ideal gas are only determined up to an arbitrary linear 
combination of the particle numbers. (This is an instance of the deeper 
problem to determine under which conditions thermodynamic variables 
are controllable; cf. the discussion in the context of Example 
\ref{ex.gibbs} below.) In practice, this gauge freedom 
(present only in the ideal gas) can be fixed by choosing a particular 
{\bf standard temperature} $T_0$ and setting arbitrarily $H_j(T_0)=0$, 
$\mu_j(T_0)=0$. 

The ideal gas law \gzit{e.ilaw} is the basis for the construction of 
gas thermometers, since the 
amount of expansion of volume in a long, thin tube can easily be read 
off from a scale along the tube. We have $V=aL$, where $a$ is the cross
section area and $L$ is the length of the filled part of the tube,
hence $T=(aP/R)L$. Thus, at constant pressure, the temperature is 
proportional to $L$. Modern thermometers work along similar principles.
\end{expl}

\section{The laws of thermodynamics}\label{s.cons}

To measure the temperature of a system,
one brings it in thermal contact with a thermometer and waits until 
equilibrium is established. ({\bf Thermal contact} is precisely the 
condition that this will happen in a short amount of time.)
The system and the thermometer will have the same temperature, 
which can be read off the thermometer. If the system is much larger
than the thermometer, this temperature will be essentially the same 
as the temperature of the system before the measurement.
The possibility of measuring temperature is sometimes called the
zeroth law of thermodynamics, following {\sc Fowler \& Guggenheim}
\cite{FowG}. For a survey of the problems involved in defining and
measuring temperature outside equilibrium, see 
{\sc Casas-V\'asquez \& Jou} \cite{CasJ}. For the history of 
temperature, see {\sc Roller} \cite{Rol} and {\sc Truesdell} \cite{Tru}.

To be able to formulate the first law of thermodynamics we need the
concept of a reversible change of states, i.e., changes
preserving the equilibrium condition. For use in later sections,
we define the concept in a slightly more general form,
writing $\alpha$ for $P$ and $\mu$ jointly.

\begin{dfn}
A {\bf state variable} is an almost everywhere continuously 
differentiable function $\phi(T,\alpha)$ defined on the
state space (or a subset of it). 
Temporal changes in a state variable that 
occur when the boundary conditions are kept fixed are called 
{\bf spontaneous changes}. 

A {\bf reversible transformation} is a continuously differentiable 
mapping 
\[
\lambda \to (T(\lambda ),\alpha(\lambda ))
\]
from a real interval into the state space; thus
$\Delta(T(\lambda ),\alpha(\lambda ))=0$. The {\bf differential}
\lbeq{3-12}
d\phi=\frac {\partial \phi} {\partial T}dT
+\frac {\partial \phi} {\partial \alpha} \cdot d\alpha 
\eeq
describes the change of a state variable $\phi$ under 
arbitrary (infinitesimal) reversible transformations. 
(Divide by $d\lambda$ and recognize the chain rule!
In formal mathematical terms, differentials are exact 1-forms on the
state space manifold.)
\end{dfn}

Reversible changes per se have nothing to do with changes in time. 
However, by sufficiently slow, quasistatic changes of the boundary 
conditions, reversible changes can often be realized approximately as 
temporal changes. The degree to which this is possible determines the 
efficiency of thermodynamic machines.

The state space is often parameterized by 
different sets of state variables, as required by the application. 
If $T=T(\kappa,\lambda)$, $\alpha=\alpha(\kappa,\lambda)$ is such a
parameterization then the state variable $g(T,\alpha)$ can be written
as a function of $(\kappa,\lambda)$,
\lbeq{e.partial0}
g(\kappa,\lambda) = g(T(\kappa,\lambda),\alpha(\kappa,\lambda)).
\eeq
This notation, while mathematically ambiguous, is common in the
literature; the names of the argument decide which function is intended.
When writing partial derivatives without arguments, this leads to
serious ambiguities. These can be resolved by writing 
$\D\Big(\frac{\partial g}{\partial \lambda}\Big)_\kappa$ for the 
partial derivative of \gzit{e.partial0} with respect to $\lambda$;
it can be evaluated using \gzit{3-12}, giving the {\bf chain rule}
\lbeq{e.partial}
\Big(\frac{\partial g}{\partial \lambda}\Big)_\kappa
=\frac{\partial g} {\partial T}
\Big(\frac{\partial T}{\partial \lambda}\Big)_\kappa
+\frac {\partial g} {\partial \alpha} \cdot
\Big(\frac{\partial \alpha}{\partial \lambda}\Big)_\kappa.
\eeq
Here the partial derivatives in the original 
parameterization by the intensive variables are written without 
parentheses.

Differentiating the equation of state \gzit{e.def}, using the chain 
rule \gzit{3-12}, and simplifying using \gzit{e.Sp} gives the 
{\bf Gibbs-Duhem equation}
\lbeq{e.GDp}
0=SdT- VdP+N\cdot d\mu
\eeq
for reversible changes. If we differentiate the 
Euler equation \gzit{e.Hp}, we obtain
\[
dH=TdS+SdT-PdV-VdP+\mu\cdot dN+N\cdot d\mu,
\]
and using \gzit{e.GDp}, this simplifies to the 
{\bf first law of thermodynamics} ({\sc Mayer} \cite{May},
{\sc Joule}\ cite{Jou},{\sc Helmholtz}\ cite{Hel}, 
{\sc Clausius} \cite{Cla})
\lbeq{3.1stp}
dH=TdS-Pd V +\mu \cdot dN.
\eeq
Considering global equilibrium from a fundamental point of view, the 
extensive variables are the variables that are conserved (or at least 
change so slowly that they may be regarded as time independent on the 
time scale of interest). In the absence of chemical reactions, the 
particle numbers, the entropy, and the Hamilton energy are conserved; 
the volume is a system size variable which, in the fundamental view, 
must be taken as infinite (thermodynamic limit) to exclude the 
unavoidable interaction with the environment. However, real systems 
are always in contact with their environment (e.g., stars through 
radiation, fluids through the forces excerted by the container),
and the conservation laws are approximate only. In thermodynamics, 
the description of the system boundary is generally reduced to the 
degrees of freedom observable at a given resolution; e.g., a heat bath 
to maintain constant temperature, or a potential well to account for 
the confinement in a container).

The result of this reduced description (for derivations, see, e.g., 
{\sc Balian} \cite{Bal}, {\sc Grabert} \cite{Gra}, 
{\sc Rau \& M\"uller} \cite{RauM}) is a dynamical effect called 
{\bf dissipation} ({\sc Thomson} \cite{Tho}), described by the 
{\bf second law of thermodynamics}
({\sc Clausius} \cite{Cla2}).
The Euler inequality \gzit{e.Hi} together with the Euler equation 
\gzit{e.Hp} only express the nondynamical part of the second law since, 
in equilibrium thermodynamics, dynamical questions are ignored: 
Axiom (ii) says that if $S,V,N$ are conserved (thermal, mechanical and 
chemical isolation) then the {\bf internal energy},
\lbeq{e.int}
U:=TS-PV+\mu\cdot N
\eeq
is minimal in equilibrium; if $T,V,N$ are conserved (mechanical and 
chemical isolation of a system at constant 
temperature $T$) then the {\bf Helmholtz energy},
\[
A:=U-TS=-PV+\mu\cdot N 
\]
is minimal in equilibrium; and if $T,P,N$ are conserved (chemical 
isolation of a system at constant temperature $T$ and pressure $P$) 
then the {\bf Gibbs energy},
\[
G:=A+PV=\mu\cdot N
\] 
is minimal in equilibrium.

The {\bf third law of thermodynamics} ({\sc Nernst} \cite{Ner}) says 
that entropy is nonnegative. 
This is a direct consequence of the monotonicity of $\Delta(T,P,\mu)$ 
and \gzit{e.Sp}.

\bigskip
{\bf Consequences of the first law.}
The first law of thermodynamics describes the observable
energy balance in a reversible process. 
The total energy flux $dH$ into the system is composed of the 
{\bf thermal energy flux} or {\bf heat flux} $TdS$,
the {\bf mechanical energy flux} $-PdV$, and the 
{\bf chemical energy flux} $\mu \cdot dN$. 

The Gibbs-Duhem equation \gzit{e.GDp} describes the energy balance 
necessary to compensate the changes $d(TS)=TdS+SdT$ of
thermal energy, $d(PV)=Pd V + V dP$ of
mechanical energy, and $d(\mu \cdot N)=\mu \cdot dN+N\cdot d\mu$ 
of chemical energy in the energy contributions to the Euler equation 
to ensure that the Euler equation
remains valid during a reversible transformation. Indeed, both
equations together imply that $d(TdS-PdV+\mu\cdot N -H)$
vanishes, which expresses the preservation of the Euler equation.

Related to the various energy fluxes are the {\bf thermal work}
\[
Q = \int T(\lambda)dS(\lambda),
\]
the {\bf mechanical work}
\[
W_\fns{mech} = -\int P(\lambda)dV(\lambda),
\]
and the {\bf chemical work}
\[
W_\fns{chem} = \int \mu(\lambda)\cdot dN(\lambda)
\]
performed in a reversible transformation. The various kinds of work 
generally depend on the path through the state space; however, the
mechanical work depends only on the end points, since the associated
process is conservative. 

As is apparent from the formulas given, thermal work is done by
changing the entropy of the system, mechanical work by changing the 
volume, and chemical work by changing the particle numbers. 
In particular, in case of thermal, mechanical, or chemical
{\bf isolation}, the corresponding fluxes vanish identically. 
Thus, constant $S$ characterizes thermally isolated, 
{\bf adiabatic} systems, constant $V$ characterizes mechanically 
isolated, {\bf closed} systems, and constant $N$ 
characterizes chemically isolated, {\bf impermeable} systems. 
(The constancy depends on all properties of a standard system: global 
equilibrium, a single phase, and the absence of chemical reactions.)
Of course, these boundary conditions are somewhat idealized situations, 
which, however, can be approximately realized in practice and are of 
immense scientific and technological importance.
(Note that the terms 'closed system' has also a 
much more general interpretation -- which we do {\em not} use in this 
paper --, namely as a conservative dynamical system.) 

The first law shows that, in appropriate units, the temperature $T$ is 
the amount of energy needed to increase in a mechanically and 
chemically isolated system the entropy $S$ by one unit. 
The pressure $P$ is, in appropriate units, the amount of energy needed 
to decrease in a thermally and chemically isolated system the volume 
$V$ by one unit. (In particular, increasing pressure decreases the 
volume; this explains the minus sign in the definition of $P$.)
The chemical potential $\mu_j$ is, in appropriate units, the amount of 
energy needed to increase in a thermally and mechanically isolated 
system the particle number $N_j$ by one. (With the traditional units, 
temperature and pressure are no longer energies.)

We see that the entropy and the volume behave just like the particle
number. This analogy can be deepened by observing that particle numbers 
are the natural measure of the amounts of ''matter'' of each kind in a 
system, and chemical energy flux is accompanied by adding or removing 
matter.
Similarly, volume is the natural measure of the amount of ''space'' a
system occupies, and mechanical energy flux in a standard system is
accompanied by adding or removing space.
Thus we may regard entropy as the natural measure of the amount of 
''heat'' contained in a system\footnote{
Thus, entropy is the modern replacement for the historical concepts of 
{\em phlogiston} and {\em caloric}, which failed to give a correct 
account of heat phenomena. 
Phlogiston turned out to be 'missing oxygen', an early 
analogue of the picture of positrons as holes, or 'missing electrons', 
in the Dirac sea. Caloric was a massless substance of heat which had 
almost the right properties, explained many effects correctly, and 
fell out of favor only after it became known that caloric could be
generated in arbitrarily large amounts from mechanical energy, thus
discrediting the idea of heat being a substance. (For the precise 
relation of entropy and caloric, see {\sc Kuhn} \cite{Kuh1,Kuh2},
{\sc Walter} \cite{Walt}, and the references quoted there.) 
In the modern picture,
the extensivity of entropy models the substance-like properties of
heat. But as there are no particles 
of space whose number is proportional to the volume, so there are 
no particles of heat whose number is proportional to the entropy.
},  
since thermal energy flux is accompanied by adding or removing heat.
Looking at other extensive quantities, we also recognize energy as the 
natural measure of the amount of ''power'', 
momentum as the natural measure of the amount of ''force'', and 
mass as the natural measure of the amount of ''inertia'' of a system. 
In each case, the notions in quotation marks are the colloquial terms 
which are associated in ordinary life with the more precise, formally 
defined physical quantities. For historical reasons, the words heat, 
power, and force are used in physics with a meaning different from the 
colloquial terms ''heat'', ''power'', and ''force''.

\bigskip
{\bf Consequences of the second law.}
The second law is centered around the impossibility of perpetual 
motion machines due to the inevitable loss of energy by  
dissipation such as friction (see, e.g., {\sc Bowden \& Leben} 
\cite{BowL}), uncontrolled radiation, etc..
This means that -- unless continually provided from the outside --
energy is lost with time until a metastable state is attained,
which usually is an equilibrium state. Therefore, the energy at
equilibrium is minimal under the circumstances dictated by the 
boundary conditions. In a purely kinematic setting as in our paper, 
the approach to equilibrium cannot be studied, and only the  
minimal energy principles -- one for each set of boundary conditions -- 
remain. 

Traditionally, the second law is often expressed in the form of
an extremal principle for some thermodynamic potential.
We derive here the extremal principles for the Hamilton energy,
the Helmholtz energy, and the Gibbs energy\footnote{
The different potentials are related by so-called Legendre transforms;
cf. {\sc Rockafellar} \cite{Roc} for the mathematical properties of 
Legendre transforms, {\sc Arnol'd} \cite{Arn} for their application in 
in mechanics, and {\sc Alberty} \cite{Alb} for their application in 
chemistry.
}, 
which give rise to the {\bf Hamilton potential}
\[
U(S,V,N) :=\max_{T,P,\mu}\,
\{TS-PV+\mu\cdot N\mid \Delta(T,P,\mu)=0;T>0;P>0\},
\]
the {\bf Helmholtz potential}
\[
A(T,V,N):=\max_{P,\mu}\,
\{-PV+\mu\cdot N\mid \Delta(T,P,\mu)=0;T>0;P>0\},
\]
and the {\bf Gibbs potential}
\[ 
G(T,P,N):=\max_\mu\,
\{\mu\cdot N\mid \Delta(T,P,\mu)=0;T>0;P>0\}.
\]
The Gibbs potential is of particular importance for everyday 
processes since the latter frequently happen at approximately constant 
temperature, pressure, and particle number. 
(For other thermodynamic potential used in practice, see 
{\sc Alberty} \cite{Alb}; for the maximum entropy 
principle, see Appendix \ref{s.maxent}.)

\begin{thm}\label{t.extstd}
{\bf (Extremal principles)}\\
(i) In an arbitrary state, 
\lbeq{e.2ndU}
H \ge U(S,V,N),
\eeq
with equality iff the state is an equilibrium state. 
The remaining thermodynamic variables are then given by
\[
T = \frac{\partial}{\partial S}U(S,V,N),~~~
P = -\frac{\partial}{\partial V}U(S,V,N),~~~
\mu = \frac{\partial}{\partial N}U(S,V,N),~~~
H = U(S,V,N).
\]
In particular, an equilibrium state is uniquely determined by 
the values of $S$, $V$, and $N$.

(ii) In an arbitrary state, 
\lbeq{e.2ndA}
H-TS \ge A(T,V,N),
\eeq
with equality iff the state is an equilibrium state.
The remaining thermodynamic variables are then given by
\[
S=-\frac{\partial A}{\partial T}(T,V,N),~~~
P=-\frac{\partial A}{\partial V}(T,V,N),~~~
\mu=\frac{\partial A}{\partial N}(T,V,N),
\]
\[
H=A(T,V,N)+TS.
\]
In particular, an equilibrium state is uniquely determined by 
the values of $T$, $V$, and $N$.

(iii) In an arbitrary state,
\lbeq{e.2ndG}
H-TS+PV \ge G(T,P,N),
\eeq
with equality iff the state is an equilibrium state.
The remaining thermodynamic variables are then given by
\[
S=-\frac{\partial G}{\partial T}(T,P,N),~~~
V=\frac{\partial G}{\partial P}(T,P,N), ~~~
\mu=\frac{\partial G}{\partial N}(T,P,N),
\]
\[
H=G(T,P,N)+TS-PV.
\]
In particular, an equilibrium state is uniquely determined by 
the values of $T$, $P$, and $N$.
\end{thm}

\bepf
We prove (ii); the other two cases are entirely similar.
\gzit{e.2ndA} and the statement about equality is a direct consequnce 
of Axiom \ref{d.phen}(ii). Thus, the difference $H-TS-A(T,V,N)$ 
takes its minimum value zero at the equilibrium value of $T$. 
Therefore, the derivative with respect to $T$ vanishes, which gives the
formula for $S$. To get the formulas for $P$ and $\mu$, we note that 
for constant $T$, the first law \gzit{3.1stp} implies
\[
dA=d(H-TS)=dH-TdS=-PdV+\mu\cdot dN.
\]
For the reversible transformation which only changes $P$ or $\mu_j$,
we conclude that $dA=-PdV$ and $dA=\mu\cdot dN$, respectively.
Solving for $P$ and $\mu_j$, respectively, implies the formulas for
$P$ and $\mu_j$.
\epf

The above results imply that one can regard each thermodynamic 
potential as a complete alternative way to describe the manifold of 
thermal states and hence all equilibrium properties.
This is very important in practice, where one usually describes
thermodynamic material properties in terms of the Helmholtz or Gibbs 
potential, using models like NRTL ({\sc Renon \& Prausnitz} \cite{RenP},
{\sc Prausnitz} et al. \cite{PraLA})
or SAFT ({\sc Chapman} et al. \cite{ChaGJR,ChaGJR2}).

The additivity of extensive quantities is reflected in corresponding 
properties of the thermodynamic potentials:

\begin{thm}\label{t.ext}
The potentials $U(S,V,N)$, $A(T,V,N)$, and $G(T,P,N)$
satisfy, for real $\lambda,\lambda^1,\lambda^2\ge 0$,
\lbeq{e.homUx}
U(\lambda S,\lambda V,\lambda N)=\lambda U(S,V,N),
\eeq
\lbeq{e.homAx}
A(T,\lambda V,\lambda N)=\lambda A(T,V,N),
\eeq
\lbeq{e.homGx}
G(T,P,\lambda N)=\lambda G(T,P,N),
\eeq
\lbeq{e.convUx}
U(\lambda^1 S^1+\lambda^2S^2,\lambda^1 V^1+\lambda^2V^2,
\lambda^1 N^1+\lambda^2N^2)
\le \lambda^1 U(S^1,V^1,N^1)+\lambda^2 U(S^2,V^2,N^2),
\eeq
\lbeq{e.convAx}
A(T,\lambda^1 V^1+\lambda^2V^2,\lambda^1 N^1+\lambda^2N^2)
\le \lambda^1 A(T,V^1,N^1)+\lambda^2 A(T,V^2,N^2),
\eeq
\lbeq{e.convGx}
G(T,P,\lambda^1 N^1+\lambda^2N^2)
\le \lambda^1 G(T,P,N^1)+\lambda^2 G(T,P,N^2).
\eeq
In particular, these potentials are convex in $S$, $V$, and $N$.
\end{thm}

\bepf
The first three equations express homogeneity and are a direct 
consequence of the definitions. Inequality \gzit{e.convAx} holds since, 
for suitable $P$ and $\mu$,
\[
\bary{lll}
A(T,\lambda^1 V^1+\lambda^2V^2,\lambda^1 N^1+\lambda^2N^2)
&=&-P(\lambda^1 V^1+\lambda^2V^2)+\mu\cdot(\lambda^1 N^1+\lambda^2N^2)\\
&=&\lambda^1(-PV^1+\mu\cdot N^1)+\lambda^2(-PV^2+\mu\cdot N^2)\\
&\le& \lambda^1 A(T,V^1,N^1)+\lambda^2 A(T,V^2,N^2);
\eary
\]
and the others follow in the same way.
Specialized to $\lambda^1+\lambda^2=1$, the inequalities express the
claimed convexity.
\epf

For a system at constant temperature $T$, pressure $P$, and particle
number $N$, consisting of
a number of parts labeled by a superscript $k$ which are separately
in equilibrium, the Gibbs energy is extensive,
\[
\bary{lll}
G&=&H-TS+PV= \D\sum H^k-T\sum S^k+P\sum V^k \\
&=& \D\sum (H^k-TS^k+PV^k)=\sum G^k.
\eary
\]
Equilibrium requires that $\sum G^k$ is minimal among all choices 
with $\sum N^k=N$, and by introducing a Lagrange multiplier vector 
$\mu^*$ for the constraints, we see that in equilibrium, the derivative 
of $\sum (G(T,P,N^k)-\mu^*\cdot N^k)$ with respect to each $N^k$ 
must vanish. This implies that 
\[
\mu^k= \frac{\partial G}{\partial N^k}(T,P,N^k)=\mu^*.
\]
Thus, in equilibrium, all $\mu^k$ must be the same.  
At constant $T$, $V$, and $N$, one can do the same argument with the 
Helmholtz potential, and at constant $S$, $V$, and $N$ with the 
Hamilton potential. In each case, the equilibrium is characterized
by the constancy of the intensive parameters.

Global equilibrium states are therefore macroscopically highly uniform. 
Their intrinsic complexity is apparent only in a microscopic treatment; 
the only macroscopic shadow of this complexity is the critical
opalescence of fluids near a critical point ({\sc Andrews} \cite{And}, 
{\sc Forster} \cite{For}).
In particular, the traditional interpretation of entropy
as a measure of disorder is often misleading\footnote{
Much more carefully argued support for this statement, with numerous
examples from teaching practice, is in {\sc Lambert} \cite{Lam}.
}: 
{\bf Macroscopic disorder} is measured by the degree to which 
macroscopic space and time correlations are absent, {\em not} by 
entropy.
In this sense, {\em global equilibrium states are the most ordered 
macroscopic states in the universe} rather than the most disordered 
ones. 

A system not in global equilibrium is characterized by macroscopic 
local inhomogeneities, indicating that the space-independent global 
equilibrium variables alone do not exhaust all available macroscopic 
information. The global equilibrium state is a state without
additional macroscopic information. The contents of the second law of
thermodynamics for global equilibrium states may therefore be phrased 
as follows (cf. Appendix \ref{s.complexity}):
{\em At global equilibrium, macroscopic order (homogeneity) is perfect 
and microscopic complexity is maximal}.

\bigskip
Using only the present axioms, one can say something about the
behavior of a system close to equilibrium, at least in the following,
idealized situation.
Suppose that a system at constant $S$, $V$, and $N$ which is close to 
equilibrium at some time $t$ reaches equilibrium at some later time 
$t^*$. Then the second law implies 
\[
0\le H(t)-H(t^*) \approx (t-t^*)\frac{dH}{dt},
\]
so that $dH/dt\le 0$. Assuming that the system is composed of two 
parts, both in equilibrium in the time interval between $t$ and $t^*$,
the time shift must be reversible on the parts, and the first law can 
be applied to them. Thus
\[
dH=\sum_{k=1,2} dH^k =\sum_{k=1,2} (T^kdS^k-P^kdV^k+\mu^k\cdot dN^k).
\]
Since $S$, $V$, and $N$ remain constant, we have $dS^1+dS^2=0$,
$dV^1+dV^2=0$, $dN^1+dN^2=0$, and since for the time shift $dH\le 0$,
we find the inequality
\[
0\ge (T^1-T^2)dS^1 - (P^1-P^2)dV^1 +(\mu^1-\mu^2)\cdot dN^1.
\]
This gives infromation about the direction of the flow in case that 
all but one of the extensive variables are known to be fixed.

In particular, at constant $V^1$ and $N^1$, we have $dS^1\le 0$ if 
$T^1>T^2$; i.e., ''heat'' (entropy) flows from the hotter part towards 
the colder part. At constant $S^1$ and $N^1$, we have $dV^1\le 0$ if 
$P^1<P^2$; i.e., ''space'' (volume) flows from lower pressure to 
higher pressure: the volume of the lower pressure part decreases and 
is compensated by a corresponding increase of the volume in the higher 
pressure part. And for a single species system at constant $S^1$ and 
$V^1$, we have $dN^1\le 0$ if $\mu^1>\mu^2$; i.e., ''matter'' (particle 
number) flows from higher chemical potential towards lower chemical 
potential. This gives temperature, pressure, and chemical potential
the familiar intuitive interpretation.

This is a shadow of the far reaching fact that, in nonequilibrium 
thermodynamics, gradients in the intensive variables 
induce a dissipative dynamics that tends to diminish these gradients,
thus enforcing (after the time needed to reach equilibrium) agreement 
of the intensive variables of different parts of a system. 
While these dynamical issues are outside the scope 
of the present work, they motivate the fact that one can control some
intensive parameters of the system by controlling the corresponding 
intensive parameters of the environment and making the walls permeable 
to the corresponding extensive quantities. This
corresponds to standard procedures familiar to everyone from ordinary 
life: heating to change the temperature, applying pressure to change 
the volume, immersion into a substance to change the chemical 
composition; also, in the more general thermal models discussed in 
Section \ref{s.detail} applying forces to displace an object, etc..

The stronger nonequilibrium version of the second law says that 
(for suitable boundary conditions) equilibrium is attained after 
some time (stictly speaking, only in the limit of infinite time).
This implies that the energy difference 
\[
\delta E=H-U(S,V,N)=H-TS-A(S,V,N)=H-TS+PV=G(S,V,N)
\]
is the amount of energy that is dissipated in order to reach 
equilibrium. In an equilibrium setting, we 
can only compare what happens to a system prepared in a nonequilibrium 
state assuming that, subsequently, the full energy difference 
$\delta E$ is dissipated so that the system ends up in an equilibrium 
state. Since few variables describe everything of interest, this 
constitutes the power of equilibrium thermodynamics. But this power is 
limited, since equilibrium thermodynamics is silent about when -- or 
whether at all -- equilibrium is reached. Indeed, in many cases, only 
metastable states are reached, which change too slowly to ever reach 
equilibrium on a human time scale.

\bigskip
The phenomenological description given so far is 
completely adequate for systems in global equilibrium with a 
single phase only and no chemical reactions.
From the formulas provided, it is now an easy step to go to various 
examples and applications.
A full discussion of global equilibrium would also involve the
equilibrium treatment of multiple phases and chemical reactions. 
However, these involve no foundational problems, and their discussion 
would offer little new compared with traditional textbook treatments. 
Hence they are not treated here.

More important is the observation that, when considered in sufficient 
detail, no physical system is truly in global equilibrium; one can 
always find smaller or larger deviations. To describe these deviations, 
extra variables are needed, resulting in a more complete but also more 
complex model. At even higher resolution, this model is again 
imperfect and an approximation to an even more complex, better model. 
This refinement process may be repeated in many, perhaps even infinitely
many stages. At the most detailed stages, we transcend the frontier of 
current knowledge in physics, but even as this frontier recedes, 
deeper and deeper stages with unknown details are imaginable.

Therefore, it is desirable to have a meta-description of thermodynamics
that, starting with a detailed model, allows to deduce the properties 
of each coarser model, in a way that all description levels are 
consistent with the current state of the art in physics. 
Moreover, the results should be as independent as possible of unknown 
details at the lowest levels.

This will be done in the remainder of the paper. First we'll look at
an anonymous bottom level, where Gibbs states are the players in the 
field and define values for arbitrary quantities.
As in this section, the intensive variables determine the state (which 
now is a more abstract object), whereas the extensive variables 
now appear as values of other abstract objects called 
quantities. This change of setting allows the natural incorporation 
of quantum mechanics, where quantities need not commute, while 
values are numbers observable in principle, hence must 
satisfy the commutative law.

Then we introduce thermal states by selecting the quantities 
whose values shall act as extensive variables in a 
thermal model. On this level, we shall be able to reproduce the
phenomenological setting of the present section from first principles;
see the discussion after Theorem \ref{t7.3}.
If the underlying detailed model is assumed to be known then the
system function, and with it all thermal properties,
are computable in principle, although we only hint at the ways to do
this numerically. We also look at a hierarchy of
thermal models based on the same bottom level description
and discuss how to decide which description levels are appropriate.

\section{Gibbs states}\label{s.gibbs}

Any fundamental description of physical systems must give account of 
the numerical values of quantities observable in experiments when the 
system under consideration is in a specified state. Moreover, the form 
and meaning of states, and what is observable in principle, must be 
clearly defined. 
For reasons given in {\sc Neumaier} \cite{Neu.ens}, we avoid using 
the customary word `observable', and consider an axiomatic conceptual 
foundation on the basis of quantities and their values. This is
consistent with the conventions adopted by the 
International System of Units (SI) \cite{SI}, who declare:
''{\em A quantity in the general sense 
is a property ascribed to phenomena, bodies, or substances that can 
be quantified for, or assigned to, a particular phenomenon, 
body, or substance. [...] 
The value of a physical quantity is the quantitative expression
of a particular physical quantity as the product of a number and a
unit, the number being its numerical value.}'' 
Since quantities can be added and multiplied, the set of all 
quantities has an algebraic structure whose properties we now formulate.

The {\bf quantities} of interest are taken to be the elements of a 
{\bf Euclidean *-algebra} $\Ez$. The simplest family of Euclidean 
*-algebras, and the only one with which the reader must be familiar 
to understand the paper, is the algebra $\Ez=\Cz^{N\times N}$ of 
square complex $N\times N$ matrices.
(This algebra models an N-level quantum system, but also quantum 
field theory in the lattice approximation.)
Here the quantites are square matrices, the constants are the multiples 
of the identity matrix, involution * is conjugate transposition, and 
the integral $\sint$ is the trace, so that all quantities are 
strongly integrable. 

For those, who want to understand a little more, we give a formally
precise definition. To say that a set $\Ez$ of quantities is a 
Euclidean *-algebra means the following:  
$\Ez$ is a complex vector space containing the complex numbers as 
{\bf constants}. 
Apart from an associative product (commutative only in the 
classical case) one has an {\bf involution} $*$ reducing on the 
constants to complex conjugation and a complex-valued {\bf integral} 
$\int$ defined on a subspace of {\bf strongly integrable} 
quantities. (The integral can often be naturally extended to a 
significantly larger class of integrable quantities.)
A {\bf partial order} $\ge$ is then defined by $g\ge 0$ iff $g^*=g$ and 
$\sint h^*gh\ge 0$ for all strongly integrable $h$, and $g\ge h$ iff 
$g-h\ge 0$.
A quantity $g$ is called  {\bf Hermitian} if $g^*=g$, and 
{\bf bounded} if $g^*g\le \alpha^2$ for some $\alpha\in\Rz$.
Apart from the standard rules for $*$-algebras 
({\sc Neumaier} \cite{Neu.ens}
discusses in detail what must be assumed and what can then be proved) 
and the linearity of the integral, we assume the following axioms
for all bounded quantities $g,h$ and all strongly integrable $h_l$:

\spd(EA1) ~ 
$g$ bounded, $h$ strongly integrable $~~\Rightarrow~~ h^*,gh,hg$ 
strongly integrable,\\
\spd(EA2) ~
$(\sint h) ^* = \sint h^*, ~~~\sint gh = \sint hg$,\\
\spd(EA3) ~
$ \sint h^* h > 0$ ~if $h \not= 0$,\\
\spd(EA4) ~
$\sint h^* g h= 0$ for all strongly integrable $h~~\Rightarrow~~ g=0$~~~
{\bf (nondegeneracy)},\\
\spd(EA5) ~
$ \sint h_l^* h_l \to 0 ~~\Rightarrow~~ \sint g h_l \to 0$,~
$\sint h_l^* g h_l \to 0$,\\
\spd(EA6) ~
$h_l\downto 0~~\Rightarrow~~ \inf\sint h_l=0$~~~
{\bf (Dini property)}.

Here, integrals extend over the longest following product or quotient
(while later, differential operators act on the shortest syntactically 
meaningful term), the monotonic limit is defined by
$g_l \downarrow 0$ iff, for every strongly integrable $h$, the sequence 
(or net) $\sint h^*g_lh$ consists of real numbers converging 
monotonically decreasing to zero. 
The reader is invited to check explicitly that (EA1)--(EA6) hold for 
$\Ez=\Cz^{N\times N}$.

Note that (EA3) implies the Cauchy-Schwarz inequality
\[
\sint (gh)^*(gh)\le \sint g^*g~\sint h^*h,
\]
which implies that every strongly integrable quantity is bounded.
Some results needed in the following are given in 
Appendix \ref{s.technical}, but proved there only in case $\Ez$ is 
finite-dimensional. To justify our treatment rigorously in case the 
algebra $\Ez$ is infinite-dimensional, one would need to use
(EA5) and (EA6) and perhaps further, technical assumptions.
However, the main text is fully rigorous if the results of Appendix 
\ref{s.technical} are assumed in addition to (EA1)--(EA6).

\begin{expls}\label{e3.1}
Apart from the matrix algebra, three basic realizations of the axioms 
are relevant to nonrelativistic physics. 
We give here some details to draw the connection with 
traditional statistical mechanics. However, the remainder is completely
independent of details how our axioms are realized; a specific 
realization is needed only when doing specific quantitative 
calculations. Therefore, the reader lacking the formal mathematical 
background to understand some details in these examples may simply 
skip them. 

(i) {\bf (Nonrelativistic classical mechanics)}
An atomic $N$-particle system is described in classical mechanics by
the phase space $\Rz^{6N}$ with six coordinates -- position 
$x^a\in\Rz^3$ and momentum $p^a\in\Rz^3$ -- for each particle.)
The algebra 
\[
\Ez_N:= C^\infty(\Rz^{6N})
\]
of infinitely differentiable, complex-valued functions 
$g(x^{1:N},p^{1:N})$  of positions and momenta
is a Euclidean *-algebra with complex conjugation as involution
and the {\bf Liouville integral}
\[
\sint g=C^{-1} \int dp^{1:N}dx^{1:N} g_N(x^{1:N},p^{1:N}),
\]
where $C$ is a positive constant.
Strongly integrable quantities are the Schwartz functions in $\Ez$.
The axioms are easily verified.

(ii) {\bf (Classical fluids)}
A fluid is classically described by an atomic system with an
indefinite number of particles. The appropriate algebra for 
a single species of monatomic particles is the 
direct sum $\Ez=\D\oplus_{N\ge 0} \Ez_N$ whose quantities are 
infinite sequences $g=(g_0,g_1,...)$ of $g_N\in\Ez_N$, with 
$\Ez_N$ as in (i), and weighted Liouville integral
\[
\sint g=\sum_{N\ge 0} 
C_N^{-1}\int dp^{1:N}dx^{1:N} g_N(x^{1:N},p^{1:N}).
\]
Here $C_N$ is a symmetry factor for the symmetry group of the
$N$-particle systen, which equals $h^{3N}N!$ for indistinguishable
particles; $h= 2\pi \hbar$ is Planck's constant.
This accounts for the Maxwell statistics and gives the correct entropy 
of mixing. Classical fluids with monatomic particles of several 
different kinds requires a tensor product of several such algebras, and 
classical fluids composed of molecules requires additional degrees
of freedom to account for the rotation and vibration of the molecules.

(iii) {\bf (Nonrelativistic quantum mechanics)} 
Let $\Hz$ be a Euclidean space, a dense subspace of a Hilbert space.
Then the algebra $\Ez:= \Lin \Hz$ of continuous linear operators 
on $\Hz$ is a Euclidean $*$-algebra with the adjoint as involution and
the {\bf quantum integral}
\[
  \sint g= \tr g,
\]
given by the trace of the quantity in the integrand.
Strongly integrable quantities are the operators $g\in\Ez$ for which 
all $hgh'$ with $h,h'\in\Ez$ are trace class. 
(This includes all linear operators of finite rank.) 
Again, the axioms are easily verified. 

\end{expls}

Our next task is to specify the formal properties of the value of a 
quantity. We assign to certain quantities $g$ (including all bounded 
quantities) a {\bf value} $\<g\>$.
Such an assignment is called a {\bf state} (of the system under 
investigation) if it has the following properties (of which
we only use the first three explicitly; the fourth would be needed 
to justify certain operations involving differentiation, 
which we silently assume to be valid):

\spd(R1)~ 
$\<1\> =1, ~~\<g^*\>=\<g\>^*,~~ \< g+h\> =\<g\> +\<h\> $,\\
\spd(R2)~ 
$\<\alpha g\> =\alpha\<g\>$ ~~~\mbox{for } $\alpha\in\Cz$, \\
\spd(R3)~ 
If $g \ge 0$ then $\<g\> \ge 0$,\\
\spd(R4)~ 
If $g_l\in\Ez,~ g_l \downarrow 0$ then $\inf_l \<g_l\> = 0$.

Note that this formal definition of a state -- always used in the 
remainder of the paper -- differs from the phenomenological version
in Section \ref{s.phen}. The connection will be made in Section 
\ref{s.eos}. 

In practice, states are assigned by well-informed and experimentally 
testable judgment concerning one's equipment. Indeed, from a practical 
point of view, {\em theory defines what an object is}: 
A gas is considered to be an ideal gas 
with certain values of temperature, pressure and volume if it behaves 
to a satisfactory degree like the mathematical model of an ideal gas
with these parameters, and a solid is considered to be a crystal with 
certain numerical properties if it behaves to a satisfactory degree like
the mathematical model of a crystal with these 
properties\footnote{\label{f.Callen}
cf. {\sc Callen} \cite[p.15]{Cal}: ``Operationally, a system is in an 
equilibrium state if its properties are consistently described by 
thermodynamic theory.''
}. 
In general, we know (or rather assume on the basis of past experience, 
claims of manufacturers, etc.) that certain materials or machines 
reliably produce states that (to a satisfactory degree for the 
purpose of the experiment or application) 
depend only on variables that are accounted for in our theory and 
that are, to a satisfactory degree, either fixed or controllable.
The nominal state of a system can be checked and, if necessary, 
corrected by {\bf calibration}, using appropriate measurements which
reveal the parameters characterizing the state.

All states arising in thermodynamics have the following
particular form.

\begin{dfn} \label{2.7.}
A {\bf Gibbs state} is defined by the values
\lbeq{2-10a}
\<g\>:=\sint e^{-S/\kbar} g, 
\eeq
where $S$, called the {\bf entropy} of the state, is a Hermitian 
quantity with strongly integrable $e^{-S/\kbar}$, satisfying the 
normalization condition
\lbeq{2-10}
\sint e^{-S/\kbar}=1,
\eeq
and $\kbar$ is the Boltzmann constant \gzit{e.kbar}.
The special case $f=S/\kbar$, $Z=1$ of Theorem \ref{2.6.} below
implies that a Gibbs state is indeed a state.
\end{dfn}

What is traditionally (and in Section \ref{s.phen}) called entropy 
and denoted by $S$
is in the present setting the value $\ol S=\<S\>$.
What is here called entropy, has in the literature on statistical 
mechanics a variety of alternative names. For example,
{\sc Gibbs} \cite{Gib} (who first noticed the rich thermodynamic 
implications of states defined by \gzit{2-10a}) called $-S$ the 
{\em index of probability};  
{\sc Alhassid \& Levine} \cite{AlhL} and {\sc Balian} \cite{Bal2}
use the name {\em surprisal} for $S$.  Our terminology is close to 
that of {\sc Mrugala} et al. \cite{MruNSS}, who call 
$S$ the {\em microscopic entropy}, and {\sc Hassan} et al. \cite{HasVL},
who call $S$ the {\em information(al) entropy operator}.

\begin{thm} \label{2.6.}~\\
(i) A Gibbs state determines its entropy uniquely.

(ii) For any Hermitian quantity $f$ with strongly integrable $e^{-f}$, 
the mapping $\<\cdot\>_f$ defined by 
\lbeq{2-6a}
\< g \>_f:=Z^{-1}\sint e^{-f} g,~~~Z:=\sint e^{-f}
\eeq
is a state. It is a Gibbs state with entropy 
\lbeq{2-8}
S:=\kbar (f+\log Z).
\eeq
(iii) The {\bf KMS condition} (cf. {\sc Kubo} \cite{Kub0},
{\sc Martin \& Schwinger} \cite{MarS})
\lbeq{e.KMS}
\<gh\>_f = \<hQ_f g\> ~~~\mbox{for bounded } g,h
\eeq
holds. Here $Q_f$ is the linear mapping defined by
\[
Q_f g :=e^{-f}ge^{f}.
\]
\end{thm}

\bepf 
(i) If the entropies $S$ and $S'$ define the same Gibbs state then 
\[
\sint (e^{-S/\kbar}-e^{-S'/\kbar}) g = \<g\>-\<g\>=0 
\]
for all $g$, hence $e^{-S/\kbar}-e^{-S'/\kbar}=0$. This implies
that $e^{-S/\kbar}=e^{-S'/\kbar}$, hence $S=S'$ by Proposition 
\ref{app2a.}.

(ii) The quantity $d:=e^{-f/2}$ is nonzero and satisfies $d^*=d$, 
$e^{-f}=d^*d\geq 0$. Hence $Z>0$ by (EA3), and $\rho:=Z^{-1}e^{-f}$ is 
Hermitian and nonnegative. For $h\ge 0$, the quantity $g=\sqrt{f}$
is Hermitian (by Proposition \ref{app2a.}) and satisfies 
$g\rho g^*=Z^{-1}(gd)(gd)^* \ge 0$, hence 
by (EA2),
\[
\<h\>_f=\<g^*g\>_f= \sint \rho g^*g =\sint g\rho g^* \ge 0. 
\] 
Moreover, $\<1\>_f =Z^{-1}\sint e^{-f}=1$. Similarly, if $g\ge 0$ then 
$g=h^*H$ with $h=\sqrt{g}=h^*$ and with $k:=e^{-f/2}h$, we get 
\[
Z\<g\>_f = \sint e^{-f}hh^*=\sint h^*e^{-f}h = \sint k^*k \ge 0.
\]
This implies (R3). the other axioms (R1)--(R4) follow easily from the 
corresponding properties of the integral. Thus $\<\cdot\>_f$ is a state.
Finally, with the definition \gzit{2-8}, we have
\[
Z^{-1}e^{-f}=e^{-f-\log Z}=e^{-S/\kbar}, 
\]
whence $\<\cdot\>_f$ is a Gibbs state.

(iii) By (EA2),
$\<hQ_fg\>_f=\sint e^{-f}hQ_fg=\sint Q_fge^{-f}h =\sint e^{-f}gh 
=\<gh\>_f$.
\epf

Note that the state \gzit{2-6a} is unaltered when $f$ is 
shifted by a constant. $Q_f$ is called the {\bf modular automorphism}
of the state $\<\cdot\>_f$ since $Q_f(gh)=Q_f(g)Q_f(h)$; for a classical
system, $Q_f$ is the identity. In the following, we shall not make use
of the KMS condition; however, it plays an important role in the 
mathematics of the thermodynamic limit (cf. {\sc Thirring} \cite{Thi}).

We call $Z$ the {\bf partition function} of $f$; it is a function of
whatever parameters appear in a particular form given to $f$ in the
applications. A large part of traditional statistical mechanics is 
concerned with the calculation of 
the partition function $Z$ and of special values when $f$ 
is given. As we shall see, much of the qualitative theory of 
statistical mechanics is completely independent of the details 
involved, and it is this part that we concentrate upon in this paper.

\begin{expl}\label{ex.canonical}
A {\bf canonical ensemble}\footnote{\label{f.ensemble} 
This traditional terminology appears to be unalterably fixed.
We shall therefore use the term {\bf ensemble} interchangable with 
state. However, we want to stress that, in the present setting, an 
ensemble may consist of a single system only, rather than -- as in the 
traditional statistical interpretation -- of a large collection 
of identically prepared systems. The latter interpretation has 
well-known difficulties to explain why each single 
macroscopic system is described correctly by thermodynamics.
}, 
is defined as a Gibbs state whose entropy is an affine function of a 
Hermitian quantity $H$, called the {\bf Hamiltonian}:
\[
S=\beta H + \const,
\]
with a constant depending on $\beta$, computable from \gzit{2-8} and
the partition function
\[
Z=\sint e^{-\beta H}
\]
of $f=\beta H$.
In particular, if $S$ is quantized then, in order that $Z$ is finite, 
$S$ and hence $H$ must have a discrete spectrum that is bounded below, 
and the partition function takes the familiar form
\lbeq{e3.3}
Z=\tr e^{-\beta H} = \sum_{n \in \cal N} e^{-\beta E_n},
\eeq
where the $E_n$ ($n\in\cal N$) are the {\bf energy levels}, the 
eigenvalues of $H$.
If the spectrum of $H$ is known, this leads to explicit formulas for 
$Z$. For example, a {\bf two level system} is defined by the energy 
levels $0,E$ (or $E_0$ and $E_0+E$, which gives the same results), 
and has
\lbeq{e.2level}
Z=1+e^{-\beta E}.
\eeq
It describes a single {\bf Fermion mode}, but also many other systems
at low temperature; cf. \gzit{e.2levelapprox}. In particular, it is the 
basis of laser-induced chemical reactions in photochemistry (see, e.g., 
{\sc Karlov} \cite{Kar}, {\sc Murov} et al. \cite{MurCH}), where 
only two molecular energy levels are activated.

For a {\bf harmonic oscillator}, defined by the energy levels $nE$, 
$n=0,1,2,\dots$ and describing a single {\bf Boson mode}, we have
\[
Z=\sum_{n=0}^\infty e^{-n\beta E} = (1-e^{-\beta E})^{-1}.
\]
Independent modes are modelled by taking tensor products of single 
mode algebras and adding their Hamiltonians, leading to spectra which 
are obtained by summing the eigenvalues of the modes in all possible 
ways. The resulting partition function is the product of the 
single-mode partition functions. From here, a thermodynamic limit 
leads to the properties of ideal gases. Then nonideal gases due to 
interactions can be handled using the cumulant expansion, as 
indicated at the end of this section. The details are outside the 
scope of this paper.
\end{expl}

Since the Hamiltonian can be any Hermitian quantity, the quantum 
partition function formula \gzit{e3.3} can in principle be used to
compute the partition function of arbitrary quantized Hermitian 
quantities.

\section{Kubo product and generating functional} \label{s.gen}

The negative logarithm of the partition function, the so-called
generating functional, plays a fundamental role in the foundation
of thermodynamics.

We first discuss a number of general properties, discovered by 
{\sc Gibbs} \cite{Gib}, {\sc Peierls} \cite{Pei}, 
{\sc Bogoliubov} \cite{Bog}, {\sc Kubo} \cite{Kub}, 
{\sc Mori} \cite{Mor}, and {\sc Griffiths} \cite{Gri}. 
The somewhat technical setting involving the Kubo inner product is
necessary to handle noncommuting quantities correctly; 
everything would be much easier in the classical case.

But the reader can be assured that (together with Appendix 
\ref{s.technical}) this section is the only technical part of 
the paper, and the effort in understanding what goes on is more than
compensated for by the ease with which everything else follows.
On a first reading, the proofs in this section may simply be skipped.

\begin{prop} Let $f$ be Hermitian such that $e^{sf}$ is strongly
integrable for all $s\in[-1,1]$. Then 
\lbeq{e.kubo}
\<g;h\>_f:=\<g E_f h\>_f,
\eeq
where $E_f$ is the linear mapping defined for Hermitian $f$ by
\[
E_f g:=\int_0^1 ds\, e^{-sf}ge^{sf},
\]
defines a bilinear, positive definite inner product 
$\<\cdot\,;\cdot\>_f$ on the algebra of quantities, 
called the {\bf Kubo} (or {\bf Mori} or {\bf Bogoliubov}) 
{\bf inner product}.
For all $f,g$, the following relations hold:
\lbeq{e.kubo2}
\<g;h\>_f^* =\<h^*;g^*\>_f.
\eeq
\lbeq{e.definit}
\<g^*;g\>_f > 0 ~~~\mbox{if } g \ne 0.
\eeq
\lbeq{e.kubo1}
\<g;h\>_f =g\<h\>_f ~~~\mbox{if $g\in \Cz$},
\eeq
\lbeq{e.kubo0}
\<g;h\>_f =\<gh\>_f ~~~\mbox{if $g$ or $h$ commutes with $f$},
\eeq
\lbeq{e.E0}
E_f g = g ~~~\mbox{if $g$ commutes with $f$},
\eeq
If $f=f(\lambda)$ depends continuously differentiably on the
real parameter vector $\lambda$ then 
\lbeq{e.deriv0}
\frac{d}{d\lambda} e^{-f} = - \Big(E_f \frac{df}{d\lambda}\Big)e^{-f}.
\eeq
\end{prop}

\bepf
(i) We have
\[
\<g;h\>_f^* =\<(gE_fh)^*\>_f = \<(E_fh)^*g^*\>_f
=\Big\<\int_0^1 ds\,e^{sf}h^*e^{-sf}g^*\Big\>_f
=\int_0^1 ds\<e^{sf}h^*e^{-sf}g^*\>_f.
\]
The integrand equals
\[
\sint e^{-f}e^{sf}h^*e^{-sf}g^* = \sint e^{sf}e^{-f}h^*e^{-sf}g^* 
=\sint e^{-f}h^*e^{-sf}g^*e^{sf} = \<h^*e^{-sf}g^*e^{sf}\>_f
\]
by (EA2), hence
\[
\<g;h\>_f^* = \int_0^1 ds\<h^*e^{-sf}g^*e^{sf}\>_f
= \Big\<h^* \int_0^1 ds\,e^{-sf}g^*e^{sf}\Big\>_f
= \<h^*E_fg^*\>_f=\<h^*;g^*\>_f.
\]
Thus \gzit{e.kubo2} holds.

(ii) Suppose that $g\ne 0$. For $s\in[0,1]$, we define $u=s/2,v=(1-s)/2$
and $g(s):= e^{-uf}ge^{vf}$. Since $f$ is Hermitian, 
$g(s)^*= e^{vf}g^*e^{-uf}$, hence by (EA2) and (EA3),
\[
\sint  e^{-f}g^*e^{-sf}ge^{sf}=\sint e^{vf}ge^{-2uf}g^*e^{vf}
=\sint g(s)^*g(s)>0, 
\]
so that
\[
\<g^*;g\>_f=\<g^*E_fg\>_f
=\int_0^1 ds\,\sint e^{-f}g^*e^{-sf}ge^{sf} > 0.
\]
This proves \gzit{e.definit}, and shows that the Kubo inner product is 
positive definite.

(iii) If $f$ and $g$ commute then $ge^{sf}=e^{sf}g$, hence 
\[
E_fg=\int_0^1 ds e^{-sf}e^{sf} g = \int_0^1 ds g = g,
\]
giving \gzit{e.E0}. The definition of the Kubo inner product then
implies \gzit{e.kubo0}, and taking $g\in\Cz$ gives \gzit{e.kubo1}.

(iv) The function $q$ on $[0,1]$ defined by
\[
q(t):= \int_0^t ds\, e^{-sf}\frac{df}{d\lambda}e^{sf}
+\Big(\frac{d}{d\lambda}e^{-tf}\Big) e^{tf}
\]
satisfies $q(0)=0$ and 
\[
\frac{d}{dt}q(t) = e^{-tf}\frac{df}{d\lambda}e^{tf}
+\Big(\frac{d}{d\lambda}e^{-tf}\Big)f e^{tf}
+\frac{d}{d\lambda}(-e^{-tf}f) e^{tf} = 0.
\]
Hence $q$ vanishes identically. In particular, $q(1)=0$, giving 
\gzit{e.deriv0}.  
\epf

As customary in thermodynamics, we use differentals to express
relations involving the differentiation by arbitrary parameters.
To write \gzit{e.deriv0} in differential form, we formally multiply by 
$d\lambda$, and obtain the {\bf quantum chain rule} for exponentials,
\lbeq{e.chain}
d e^{-f} = (- E_fd f) e^{-f}.
\eeq
If the $f(\lambda)$ commute for all values of $\lambda$
then the quantum chain rule reduces to the classical chain rule.
Indeed, then $f$ commutes also with $\frac{df}{d\lambda}$; hence 
$E_f\frac{df}{d\lambda} = \frac{df}{d\lambda}$, and $E_fd f = df$.

\bigskip
{\em The following theorem is central to the mathematics of 
thermodynamics.}
As will be apparent from the subsequent discussion, part (i) is the 
abstract mathematical form of the second law of thermodynamics, 
part (iii) is the abstract form of the first law, and part (ii)
allows the actual computation of thermal properties from
microscopic assumptions.

\begin{thm} \label{t3.3}
Let $f$ be Hermitian such that $e^{sf}$ is strongly
integrable for all $s\in[-1,1]$. 

(i) The {\bf generating functional}
\lbeq{e.gen}
W(f):=- \log \sint e^{-f}
\eeq
is a concave function of the Hermitian quantity $f$.
In particular,
\lbeq{e.GB}
W(g) \le W(f)+\<g-f\>_f.~~~\mbox{\bf (Gibbs-Bogoliubov inequality)} 
\eeq
Equality holds in \gzit{e.GB} iff $f$ and $g$ differ by a constant. 

(ii) For Hermitian $g$, the {\bf cumulant expansion}
\lbeq{e.cumulant}
W(f+\tau g)
= W(f)+\tau\<g\>_f - \frac{\tau^2}{2}(\<g\>_f^2-\<g;g\>_f) + O(\tau^3)
\eeq
holds if the coefficients are finite.

(iii) If $f=f(\lambda)$ and $g=g(\lambda)$ depend continuously 
differentiably on $\lambda$ then the following {\bf differentiation 
formulas} hold:
\lbeq{e.diff}
d\<g\>_f = \<dg\>_f-\<g;df\>_f+\<g\>_f\<df\>_f, 
\eeq
\lbeq{e.diffW}
dW(f)=\<df\>_f.
\eeq
(iv) The entropy of the state $\<\cdot\>_f$ is
\lbeq{e.ent}
S=\kbar(f-W(f)).
\eeq
\end{thm}

\bepf
We prove the assertions in reverse order.

(iv) Equation \gzit{e.gen} says that $W(f)=-\log Z$, which together 
with \gzit{2-8} gives \gzit{e.ent}.

(iii) We have 
\[
\bary{lll}
d\sint ge^{-f} &=& \sint dg e^{-f} + \sint gde^{-f}
=\sint dge^{-f}-\sint gE_fd fe^{-f}\\
&=&\sint(dg-gE_fd f)e^{-f} = Z\<dg-gE_fd f\>_f.
\eary
\]
On the other hand, 
$d\sint ge^{-f} = d(Z\<g\>_f)=dZ\<g\>_f+Zd\<g\>_f$, so that
\lbeq{e.s1}
dZ\<g\>_f+Zd\<g\>_f = Z\<dg-gE_fd f\>_f = Z\<dg\>_f-Z\<g;df\>_f.
\eeq
In particular, for $g=1$ we find by \gzit{e.kubo1} that 
$dZ=-Z\<1;df\>_f=-Z\<df\>_f$. Now \gzit{e.diffW} follows from
$dW(f)=-d\log Z =-dZ/Z = \<df\>_f$, and solving \gzit{e.s1} for 
$d\<g\>_f$ gives \gzit{e.diff}.

(ii) We introduce the function $\phi$ defined by
\[
\phi(\tau):=W(f+\tau g),
\]
From \gzit{e.diffW}, we find $\phi'(\tau) = \<g\>_{f+\tau g}$
for $f,g$ independent of $\tau$, and by differentiating this again,
\[
\phi''(\tau)=\D\frac{d}{d\tau}  \<g\>_{f+\tau g}
=\D-\Big\<g\frac{E_fd (f+\tau g)}{d\tau}\Big\>_{f+\tau g}
+\<g\>_{f+\tau g}^2.
\]
In particular, 
\lbeq{e.x5}
\phi'(0) = \<g\>_f,~~~
\phi''(0) = \<g\>_f^2-\<gE_f g\>_f= \<g\>_f^2-\<g;g\>_f.
\eeq
A Taylor expansion now implies \gzit{e.cumulant}. 

(i) Since the Cauchy-Schwarz equation 
for the Kubo inner product implies 
\[
\<g\>_f^2=\<g;1\>_f^2\le \<g;g\>_f\<1;1\>_f= \<g;g\>_f, 
\]
we see that 
\[
\frac{d^2}{d\tau^2} W(f+\tau g)\Big|_{\tau=0}\le 0
\]
for all $f,g$. This implies that $W(f)$ is concave.
Moreover, replacing $f$ by $f+sg$, we find that $\phi''(s)\le 0$ for
all $s$. The remainder form of Taylor's theorem therefore gives
\[
\phi(\tau)=\phi(0)+\tau\phi'(0)+\int_0^\tau ds (\tau-s)\phi''(s)
\le \phi(0)+\tau\phi'(0),
\]
and for $\tau=1$ we get
\lbeq{e.x6}
W(f+g)\le W(f)+\<g\>_f.
\eeq
\gzit{e.GB} follows for $\tau=1$ upon replacing $g$ by $g-f$.

By the derivation, equality holds in \gzit{e.x6} only if $\phi''(s)=0$ 
for all $0<s<1$. By \gzit{e.x5}, applied with $f+sg$ in place of $f$, 
this
implies $\<g\>_{f+sg}^2 = \<g;g\>_{f+sg}$. Thus we have equality in 
the Cauchy-Schwarz argument, forcing $g$ to be a multiple of $1$.
Therefore equality in the Gibbs-Bogoliubov inequality \gzit{e.GB} 
is possible only if $g-f$ is a constant.
\epf

\begin{thm} \label{t4.5} 
Let $S_c$ be the entropy of a reference state. Then, for an arbitrary 
Gibbs state $\<\cdot\>$ with entropy $S$,
\lbeq{e4.5}
\< S \> \le \< S_c\>,
\eeq
with equality only if $S_c =S$.
\end{thm}

\bepf
Let $f=S/\kbar$ and $g=S_c/\kbar$. Since $S$ and $S_c$ are 
entropies, $W(f)=W(g)=0$, and the Gibbs-Bogoliubov inequality 
\gzit{e.GB} gives $0\le \<g-f\>_f = \<S_c-S\>/\kbar$.
This implies \gzit{e4.5}. If equality holds then equality holds in 
\gzit{e.GB}, so that $S_c$ and $S$ differ only by a constant.
But this constant vanishes since the values agree.
\epf

The difference
\lbeq{4-5}
\< S_c-S\>  =\< S_c\> -\< S\>  \ge 0
\eeq
is known as {\bf relative entropy}.
It can be interpreted as the amount of information
in a state $\< \cdot \>$ which cannot be explained by
the reference state. This interpretation makes sense since 
the relative entropy vanishes precisely for the reference state. 
A large relative entropy therefore indicates that the state contains 
some important information not present in the reference state.

\bigskip
{\bf Approximations.} 
The cumulant expansion is the basis of a well-known
approximation method in statistical mechanics. Starting from special
reference states $\<\cdot\>_f$ with explicitly known $W(f)$ and $E_f$ 
(corresponding to so-called explicitly solvable models), one obtains 
inductively expressions for values in these states by 
applying the differentiation rules. (In the most important cases,
the resulting formulas for the values are commonly
referred to as a {\bf Wick theorem}, cf. {\sc Wick} \cite{Wic}. 
For details, see textbooks on statistical mechanics, 
e.g., {\sc Huang} \cite{Hua}, {\sc Reichl} \cite{Rei}.)

From these, one can calculate the coefficents in the cumulant 
expansion; note that higher order terms can be found 
by proceeding as in the proof, using further differentiation. 
This gives approximate generating functions (and by 
differentiation associated values) for Gibbs states 
with an entropy close to the explicitly solvable reference state.
From the resulting generating function and the differentiation 
formulas \gzit{e.diff}--\gzit{e.diffW},
one gets as before the values for the given state.

The best reference state  $\<\cdot\>_f$ to use for a given Gibbs state 
$\<\cdot\>_g$ can be obtained by minimizing the upper bound in the
Gibbs-Bogoliubov inequality \gzit{e.GB} over the $f$ for which an 
explicit generating function is known.
Frequently, one simply approximates $W(g)$ by the minimum of this
upper bound,
\lbeq{e.meanfield}
W(g) \approx W_m(g):=\inf_f \Big(W(f)+\<g-f\>_f\Big).
\eeq
Using $W_m(g)$ in place of $W(g)$ defines a so-called 
{\bf mean field theory}; cf. {\sc Callen} \cite{Cal}.
For computations from first principles (quantum field theory), see, 
e.g., the survey by {\sc Berges} et al. \cite{BerTW}.

\section{Limit resolution and statistics} \label{s.limit}

In statistical mechanics, our values of quantities are called 
expectations, and refer to the mean over an ensemble of (real or 
imagined) identically prepared systems. While we keep the notation 
with pointed brackets familiar from statistical mechanics, we want to 
avoid any probabilistic connotation, hence use our more neutral term 
vthat does not require a statistical interpretation.

We now discuss in which sense the interpretation of the values of
quantities as objective, observer-independent properties is valid.
The key is an analysis of the uncertainty inherent in the
description of a system by a state, based on the following result.
(Cf. {\sc Neumaier} \cite{Neu.ens} for a preliminary version of this
interpretation.) 

\begin{prop} 
For Hermitian $g$, 
\lbeq{e.res0}
\<g\>^2 \le \<g^2\>.
\eeq
Equality holds if $g=\<g\>$. 
\end{prop}

\bepf
Put $\ol g = \<g\>$. Then $0\le\<(g-\ol g)^2\>=\<g^2-2\ol g g+\ol g^2\>
=\<g^2\>-2\ol g \<g\>+\ol g^2=\<g^2\>-\<g\>^2$.
This gives \gzit{e.res0}. If $g=\ol g$ then equality holds in this
argument.
\epf

As a consequence, we can define the {\bf limit resolution}
\lbeq{e.res}
\res(g):=\sqrt{\<g^2\>/\<g\>^2-1},
\eeq
of a Hermitian quantity $g$ with nonzero value $\<g\>$, 
an uncertainty measure specifying how accurately one can treat $g$ 
as a sharp number, given by this value. 

In experimental practice, the limit resolution is a lower bound 
on the relative accuracy with which one can expect $\<g\>$ to be 
determinable reliably\footnote{
The situation is analogous to the limit resolution with which one can
determine the longitude and latitude of a city such as Vienna.
Clearly these are well-defined only up to some limit resolution
defined by the diameter of the city.
}\ 
from measurements of a single system at a single 
time. In particular, a quantity $g$ is considered to be 
{\bf significant} if $\res(g)\ll 1$, while it is {\bf noise} if 
$\res(g)\gg 1$. If $g$ is a quantity and $\widetilde g$ is a good 
approximation of its value then $\eps:=g-\widetilde g$ is 
noise. 

For a single system at a single time, values $\<g\>$ have in 
intermediate cases, where $\res(g)$ has the order of unity, no 
experimentally testable meaning. However, time averages of $\ol g(t_l)$
at different times $t_l$ may have a meaning -- they are the 
values of the quantity $\tilde g$ defined as the average of the 
$g(t_l)$, and by the law of large numbers, this quantity may be 
significant even if each single $g(t_l)$ is not.

Thus, this terminology captures correctly the experimental practice, 
without imposing any statistical or probabilistic connotations. 
On the contrary, it determines the precise conditions under which  
statistical reasoning is justified.

\bigskip
{\bf Objective probability.}
The exposition in {\sc Whittle} \cite{Whi} (or, in more abstract 
terms, already {\sc Gelfand \& Naimark} \cite{GelN}) shows that, 
if $H$ and the $X_j$ are pairwise commuting, there is a way consistent 
with the traditional axioms for probability theory (as formalized by 
{\sc Kolmogorov} \cite{Kol}) to define, 
for any Gibbs state in our sense, random 
variables $H$ and $X_j$ such that the expectation of all sufficiently
regular functions $f(H,X)$ defined on the joint spectrum of $(H,X)$
agrees with the value of $f$. Thus, in the pairwise commuting 
case, it is always possible to construct a probability interpretation 
for the quantities, 
{\em completely independent of any assumed microscopic reality}. 

However, if -- as in quantum systems -- the extensive quantities do 
not commute, a probabilistic interpretation in the traditional 
Kolmogorov sense is no longer possible. 
Insisting in the quantum case on a probabilistic 
interpretation brings with it the well-known difficulties of quantum 
philosophy for systems which cannot be repeatedly prepared.
It is gratifying that the present setting avoids these difficulties in 
an elegant way: There is no need at all to introduce probabilities 
into the description; everything applies to the single system under 
study.

\bigskip
{\bf Subjective probability.}
Our formalism is closely related to that used in statistics for random 
phenomena expressible in terms of exponential families; cf. Remark 
\ref{r7.2}(v) below. In such a context, a Bayesian, subjective 
probability interpretation (see, e.g., {\sc Barndorff-Nielsen} 
\cite{BarN,BarN2}) of the formalism is possible; 
then our integral defines the noninformative prior, i.e., the least 
informative ensemble. (The noninformative prior is typically improper, 
i.e., not a probability distribution, since $\sint 1$ is usually not 
defined). 

While the Bayesian view gives, in my opinion, a misleading picture of 
statistical mechanics (see Appendix \ref{s.maxent}), such an 
interpretation is formally allowed (cf. Appendix \ref{s.complexity})
if -- and only if -- 
(i) {\em correct, complete and exact information about the expectation 
of all relevant quantities}\  is assumed to be known, and 
(ii) the noninformative prior is fixed by the above constructions, 
namely as the correctly weighted Liouville measure in classical physics
and as the microcanonical ensemble in quantum physics. 
Only this guarantees that the 'knowledge' assumed and hence the 
results obtained are completely impersonal and objective.
However, this kind of knowledge is clearly completely hypothetical 
and has nothing to do with the real, subjective knowledge of real 
observers. Thus, it is better to avoid any reference to knowledge; 
it is neither useful nor necessary but confuses the subject.

\bigskip
{\bf Statistics.}
In a context where many repeated experiments are feasible, 
states can be given a frequentist interpretation, where $\<g\>$ is the 
expectation of $g$, empirically defined as an average over 
many realizations. 
In this case, $\res(g)$ becomes the standard deviation of $g$, divided 
by the absolute value of the expectation; therefore, it measures the 
relative accuracy of the realizations. 
However, as discussed in detail in the 
excellent survey by {\sc Sklar} \cite{Skl}, this interpretation
has significant foundational problems, already in the framework of 
classical physics. And it seems especially
inappropriate in equilibrium thermodynamics, where a tiny number 
of macroscopic observations on a single system completely determine 
its state to engineering accuracy.

In the following, we shall, therefore, completely avoid any reference 
to probability and statistics. While keeping the terminology of 
ensembles in the cases where it is established (the canonical and the 
grand canonical ensemble), we use the term 
ensemble$^{\mbox{\protect\ref{f.ensemble}}}$ in a non-stochastic manner.
We denote by it just a way of defining a state which provides
for all quantities objective values satisfying the above properties.
{\em In contrast to tradition, we take $\<\cdot\>$ {\rm not} to have an 
intrinsic probabilistic meaning as an average over a large statistical 
ensemble of `true' microstates.}

In this way, we delegate statistics to its role as {\em the art of 
interpreting measurements}, as in classical physics. Indeed, to have
a consistent interpretation, real experiments must be designed such that
they allow one to determine approximately the properties of the state 
under study, hence the values of all quantities of interest.
The uncertainties in the experiments imply approximations, which, 
if treated probabilistically, need an {\em additional} probabilistic 
layer accounting for measurement errors. 
Expectations from this secondary layer, which involve probabilistic
 statements about situations that are uncertain due to neglected 
but observable details (cf. {\sc Peres} \cite{Per}), happen to have the 
same formal properties as the values on the primary layer 
(which contain a complete description of what is observable at a given 
energy scale), though their physical origin and meaning is completely 
different.

In the following, we only consider the primary level, where measurement
errors are considered to be absent and an objective idealized 
description is assumed to be valid.

\section{The zeroth law: Thermal states} 
\label{zeroth}

We now restrict our attention to a restricted but very important class 
of Gibbs states, those describing thermal states.

In thermodynamics, one distinguishes between extensive
and intensive variables. {\em Extensive variables} such as mass, charge,
or volume depend additively on the size of the system; the global
quantities are given by fields interpreted as densities integrated 
over the region occupied by the system. 
The conjugate {\em intensive variables} are given by fields 
interpreted as field strengths; they act as parameters defining the 
state. They cannot be viewed as densities; instead, their differences 
or gradients have physical significance as the sources for 
thermodynamic forces. In particular, in the single-phase global 
equilibrium case treated in Section \ref{s.phen}, where the densities 
are constant, the extensive variables are densities multiplied by the 
system size, hence scale linearly with the size of the system, while 
intensive variables are invariant under a change of system size. 
(We do {\em not} use the alternative convention to call extensive any
variable which scales linearly with the system size, and intensive
any variable invariant under a change of system size.)

Thermal states are good models for macroscopic physical 
systems that are homogeneous on the level used for modeling. 
Indeed, as we shall see, they have all the properties traditionally 
postulated in thermodynamics. We distinguish four nested levels of 
thermal descriptions, {\em global, local, microlocal} and {\em quantum 
equilibrium}. The highest and computationally simplest level, 
global equilibrium, is concerned with macroscopic situations 
described by finitely many space- and time-independent variables. 
The next level, local equilibrium, treats macroscopic situations in a 
hydrodynamical or continuum mechanical description, where the relevant 
variables are finitely many space- and time-dependent fields, 
though for stirred chemical reactions, the space-dependence can be 
ignored. The next deeper level, microlocal\footnote{
The term microlocal for a phase space dependent analysis is taken
from the literature on partial differential equations; see, e.g., 
{\sc Martinez} \cite{Mar}.
}~ 
 equilibrium, treats mesoscopic situations 
in a kinetic description, where the relevant variables are finitely 
many fields depending on time, position, and momentum; 
cf. {\sc Balian} \cite{Bal2}. The bottom level is the 
microscopic regime, where we must consider quantum equilibrium,
which is described in terms of quantum dynamical semigroups.
The relations between the different description levels will be 
discussed in Section \ref{s.model}. Apart from descriptions on these 
clear-cut levels, there are also various hybrid descriptions, where 
the most important part of a system is described on a more detailed 
level than the remaining parts. 

In global equilibrium, all thermal variables are constant 
throughout the system, except at phase boundaries, where the extensive 
variables may exhibit jumps and only the intensive variables remain 
constant. This is sometimes referred to as the zeroth law of 
thermodynamics and characterizes global equilibrium; it allows one to
measure intensive variables (like temperature) by bringing a
calibrated instrument that is sensitive to this variable
(for temperature a thermometer) into equilibrium with the system to be 
measured. For local or microlocal equilibrium, the same intuition 
applies, but with fields in place of variables. Then extensive 
variables are densities represented by distributions that can be 
meaningfully integrated over bounded regions, whereas intensive 
variables are nonsingular fields (e.g., pressure) whose integrals 
are physically irrelevant.

Although dynamics is important for systems not in global equilibrium, 
we ignore dynamical 
issues completely in the remainder of this paper.  
We take a strictly kinematic point of view, and look only at a 
single phase without chemical reactions.
In a future paper, we shall extend the present setting to cover the 
dynamics of the nonequilibrium case and deduce quantitatively the 
dynamical laws ({\sc Beris \& Eswards} \cite{BerE}, 
{\sc Oettinger} \cite{Oet}) from microscopic properties, including 
phase formation, chemical reactions, and the approach to equilibrium; 
cf. {\sc Balian} \cite{Bal2}, {\sc Grabert} \cite{Gra}, 
{\sc Rau \& M\"uller} \cite{RauM}, {\sc Spohn} \cite{Spo}.

\bigskip
In our present setting, the intensive variables are, as in 
Section \ref{s.phen}, numbers 
characterizing certain Gibbs states, parameterizing the entropy. 
To each admissible combination of intensive variables there is a 
unique thermal state providing values for all quantities. 
The extensive variables then appear as the values of 
corresponding extensive quantities. 

A basic extensive quantity present in each thermal system is
the {\bf Hamilton energy} $H$; in addition, there are further basic
extensive quantities which we call $X_j$ ($j\in J$) and collect in a
vector $X$, indexed by $J$. The number and meaning of the extensive
variables depends on the type of the system; typical examples are
given in Table \ref{3.t.} in Section \ref{s.model}.

In the context of statistical mechanics (cf. Examples \ref{e3.1}), 
the Euclidean *-algebra $\Ez$ 
is an algebra of functions (for classical physics) or linear operators 
(for quantum physics), and $H$ is a particular function or linear 
operator called the {\bf Hamiltonian}; 
specifying the precise form of the Hamiltonian is 
essentially equivalent to specifying the system under consideration.
The form of the operators $X_j$ depends on the level of thermal 
modeling; for further discussion, see Section \ref{s.model}.

For qualitative theory and for deriving semiempirical 
recipes, there is no need to know details about $H$ or $X_j$; 
it suffices to treat them as primitive objects. The advantage we gain 
from our less detailed setting is that a much simpler machinery 
than that of statistical mechanics proper suffices to reconstruct all 
of {\em phenomenological} thermodynamics. 

It is intuitively clear from the above informal definition 
of extensive variables that the only functions of
independent extensive variables that are again extensive can be
linear combinations, and it is a little surprising that {\em the whole
machinery of equilibrium thermodynamics follows from a formal version 
of the simple assumption that in thermal states the entropy is
extensive}. We take this to be the mathematical expression of the 
zeroth law and formalize this assumption in a precise mathematical
definition. 

For notational simplicity, we consider mainly the case of global 
equilibrium, where there are only finitely many extensive variables. 
Everything extends, however, with (formally trivial but from a 
rigorous mathematical view nontrivial) changes to local and 
microlocal equilibrium, where extensive variables are fields, 
provided the sums are replaced by appropriate integrals;
cf. {\sc Oettinger} \cite{Oet}.

\begin{dfn} \label{3.1.}
A {\bf thermal system} is defined by a family of Hermitian 
{\bf extensive variables} $H$ and $X_j$ ($j\in J$) from a Euclidean 
*-algebra. A {\bf thermal state} of a thermal system is 
a Gibbs state whose entropy $S$ is a linear combination of the 
basic extensive quantities of the form
\lbeq{3-2}
S=T^{-1}\Big(H-\sum _{j\in J}\alpha_jX_j\Big) 
=T^{-1}(H-\alpha\cdot X) ~~~
\mbox{\bf (zeroth law of thermodynamics)} 
\eeq
with suitable real numbers $T\not=0$ and $\alpha_j$ ($j\in J$).
(Here $\alpha$ and $X$ are the vectors with components $\alpha_j$ 
($j\in J$) and $X_j$ ($j\in J$, respectively.)
Thus the value of an arbitrary quantity $g$ is
\lbeq{e.thermal}
\ol g:=\<g\>=\sint e^{-\beta(H-\alpha\cdot X)}g,
\eeq
where 
\lbeq{3-8}
\beta =\frac {1} {\kbar T}.
\eeq
\end{dfn}

The numbers $\alpha_j$ are called the {\bf intensive
variables conjugate to} $X_j$, the number $T$ is called the
{\bf temperature}, and $\beta$ the {\bf coldness}.
$\ol S,\ol H,\ol X,T$, and $\alpha$ are called the
{\bf thermal variables} of the system.
Note that the extensive variables of traditional thermodynamics are 
in the present setting not represented by the extensive quantities 
$S,H,X_j$ themselves but by their values $\ol S,\ol H,\ol X$.

Since we can write the zeroth law (\ref{3-2}) in the form
\lbeq{e.euler}
H=TS+\alpha \cdot X,
\eeq
called the {\bf Euler equation}, the temperature $T$ is considered to 
be the intensive variable conjugate to the entropy $S$. 

As already indicated in Example \ref{ex.ideal}, measuring intensive 
variables is based upon the empirical fact (which cannot be formulated 
precisely in our purely kinematic setting) that two systems 
in close contact which allows free exchange of some extensive quantity
tend to relax to a joint equilibrium state, in which the corresponding 
intensive variable is the same in both systems. If a small measuring 
device is brought into close contact with a large system, the joint 
equilibrium state will be only minimally different from the original 
state of the large system; hence the intensive variables of the 
measuring device will move to the values of the intensive variables 
of the large system in the location of the measuring device. This 
allows to read off their value from a calibrated scale.

Many treatises of equilibrium thermodynamics take this possibility of
measuring temperature to be the contents of the zeroth law of 
thermodynamics. The present, different choice for the  zeroth law is
superior since it has a definite mathematical content which has far 
reaching consequences.
Indeed, as we shall see, our definition already implies the first and 
second law, and (together with a quantization condition) the third law 
of thermodynamics.

\begin{rems}\label{r7.2} 
(i) We emphasize that the extensive quantities $H$ and $X_j$ are 
independent of the intensive quantities $T$ and $\alpha$, while $S$, 
defined by \gzit{3-2}, is an extensive quantity defined only when 
values for the intensive quantities are prescribed.
Note that $T$ and $\alpha$ vary with the state, and the entropy 
depends on them via (\ref{3-2}); hence values also depend 
on the particular state a system is in. It is crucial to distinguish 
between the quantities $H$ or $X_j$ (which are independent of $T$ 
and $\alpha$ and hence depend on the system but not on the state) 
and their values $\ol{H}=\< H \>$ or $\ol{X}_j=\< X_j \>$ 
(which change when the state of the system changes). 

(ii) Although in thermodynamics, the emphasis is on the values of 
the thermal variables, it is important to realize that a thermal 
state gives complete information about the values 
\gzit{e.thermal} of {\em arbitrary} quantities, not only the extensive 
ones. The few numbers or fields specifying $T$ and $\alpha$
fully characterize a thermal state of a given system, while a general 
non-thermal state usually has a vastly higher complexity. 
In particular, global equilibrium is characterized by a small, finite
number of variables, while already local equilibrium involves
infinitely many degrees of freedom.

(iii) Of course, the number of parameters depends on the true physical 
situation. A system in local equilibrium only cannot be adequately 
described by the few variables characterizing global equilibrium. 
The question of selecting the right set of extensive quantities for 
an adequate description is discussed in Section \ref{s.model}.

(iv) An arbitrary linear combination
\lbeq{3-6}
S=\gamma H+h_0X_0+\dots +h_sX_s
\eeq
can be written in the form \gzit{3-2} with $T=1/\gamma $ and 
$\alpha_j=-h_j/\gamma $, provided that $\gamma \not=0$; and indeed, 
\gzit{3-6} is the mathematically more natural form, which also allows 
states of infinite temperature that are excluded in \gzit{3-2}. 
However, the formulation as \gzit{3-2} seems to be unalterably fixed by 
tradition, so we shall use it here, too.
This shows that the coldness $\beta$ is a more natural variable than 
the temperature $T$; it figures prominently in statistical mechanics. 
In the limit $T \to 0$, a system becomes infinitely cold, giving 
intuition for the unattainability of zero absolute temperature. 
States of negative temperature, i.e., negative coldness, must 
therefore be considered to be hotter, i.e., less cold, than states of 
any positive temperature, 
cf. {\sc Landau \& Lifschitz} \cite{LanL}. To model thermal 
behavior at negative or infinite temperature, one can alternatively 
introduce a dummy temperature $\widehat T=1$, a 
dummy Hamiltonian $\widehat H=0$, and treating the true Hamiltonian 
and the coldness as additional components of $X$ and $\alpha$, 
respectively. 

(v) In mathematical statistics, there is a large body of work on
{\em exponential families}, 
which is essentially the mathematical equivalent 
of the concept of a thermal state over a commutative algebra; 
see, e.g., {\sc Barndorff-Nielsen} \cite{BarN}.
In this context, the values of the extensive quantities 
define a {\em sufficient statistic}, from which the whole distribution
can be reconstructed (cf. Theorem \ref{3.3.} below 
and the remarks on objctive probability in Section \ref{s.limit}). 
This is one of the reasons why
exponential families provide a powerful machinery for statistical 
inference; see, e.g., {\sc Bernardo \& Smith} \cite{BerS}. 
For recent extensions to quantum statistical inference, see, e.g.,  
{\sc Barndorff-Nielsen} et al. \cite{BarN2} and the references there.
\end{rems}

\section{The equation of state}\label{s.eos}

Not every combination $(T,\alpha)$ of intensive variables defines
a thermal state; the requirement that $\<1\>=1$ enforces a restriction
of $(T,\alpha)$ to a manifold of admissible thermal states.

\begin{thm} \label{t.eos}

Suppose that $T>0$.

(i) For any $\kappa>0$, the {\bf system function} $\Delta$ defined by
\lbeq{e.eosa}
\Delta(T,\alpha):= \kappa T \log \sint e^{-\beta (H-\alpha\cdot X)}
\eeq
is a convex function of $T$ and $\alpha$. It vanishes only if $T$
and $\alpha$ are the intensive variables of a thermal state.

(ii) In a thermal state, the intensive variables are related by the 
{\bf equation of state}
\lbeq{e.eos}
\Delta(T,\alpha)=0.
\eeq
The {\bf state space} is the set of $(T,\alpha)$ satisfying 
\gzit{e.eos}.

(iii) The values of the extensive variables are given by
\lbeq{e.Sx}
\ol S=\Omega\frac{\partial \Delta}{\partial T}(T,\alpha),~~~
\ol X=\Omega\frac{\partial \Delta}{\partial \alpha}(T,\alpha)~~~
\mbox{for some } \Omega>0,
\eeq
and the {\bf phenomenological Euler equation}
\lbeq{e.H}
\ol H=T \ol S + \alpha \cdot \ol X.
\eeq

(iv) Regarding $\ol S$ and $\ol X$ as functions of $T$ and $\alpha$, 
the matrix
\lbeq{7-12}
\Sigma:= 
\left(\begin{array}{cc}
\D\frac{\partial\ol{S}}{\partial T}& 
\D\frac{\partial\ol{S}}{\partial\alpha } \\
 ~\\
\D\frac{\partial\ol{X}}{\partial T}& 
\D\frac{\partial\ol{X}}{\partial\alpha }
\end{array}\right)
\eeq
is symmetric and positive semidefinite; in particular, we have the 
{\bf Maxwell reciprocity relations}
\lbeq{7-10}
\frac {\partial\ol{X}_i} {\partial\alpha_j}
=\frac {\partial\ol{X}_j} {\partial\alpha_i},\quad
\frac {\partial\ol{X}_i} {\partial T}
=\frac {\partial\ol{S}} {\partial\alpha_i},
\eeq
and the {\bf stability conditions}
\lbeq{7-13}
\frac {\partial\ol{S}} {\partial T}\ge 0,~~ 
\frac {\partial\ol{X}_j} {\partial\alpha_j}\ge 0~~~(j\in J).
\eeq
\end{thm}

\bepf
By Theorem \ref{t3.3}(i), the function $\phi$ defined by
\[
\phi(\alpha_0,\alpha):=\log\sint e^{-(\alpha_0H-\alpha\cdot X)} 
= -W(\alpha_0H-\alpha\cdot X)
\]
is a convex function of $\alpha_0$ and $\alpha$. Put 
$\Omega=\kbar/\kappa$. Then, by Proposition \ref{e.convex}, 
\lbeq{e.co}
\Delta(T,\alpha)=-\kappa T W(\beta(H-\alpha\cdot X))
= \kappa T \phi\Big(\frac{1}{\kbar T},\frac{\alpha}{\kbar T}\Big)
\eeq
is also convex. The condition $\Delta(T,\alpha)=0$ is equivalent to
\[
\sint e^{-S/\kbar} = \sint e^{-\beta(H-\alpha\cdot X)} 
=e^{\Delta/\kappa T} =1,
\]
the condition for a thermal state. This proves (i) and (ii).

(iii) The formulas for $\ol S$ and $\ol X$ follow by differentiation of 
\gzit{e.co} with respect to $T$ and $\alpha$, using \gzit{e.diffW}.
Equation \gzit{e.H} follows by taking values in 
\gzit{e.euler}, noting that $T$ and $\alpha$ are real numbers. 

(iv) By (iii), the matrix 
\[
\Sigma= 
\left(\begin{array}{cc}
\D\frac{\partial^2\Delta}{\partial T^2}& 
\D\frac{\partial^2\Delta}{\partial T\partial\alpha} \\
 ~\\
\D\frac{\partial^2\Delta}{\partial\alpha \partial T}& 
\D\frac{\partial^2\Delta}{\partial\alpha^2}
\end{array}\right)
\]
is the Hessian matrix of the convex function $\Delta$. Hence $\Sigma$ 
is symmetric and positive semidefinite. \gzit{7-10} expresses the 
symmetry of $\Sigma$, and \gzit{7-13} holds since the diagonal entries
of a  positive semidefinite matrix are nonnegative.
\epf

\begin{rems}\label{r5.2}
(i) For $T<0$, the same results hold, with the change that $\Delta$ is 
concave instead of convex, $\Sigma$ is negative semidefinite, and
the inequality signs in \gzit{7-13} are reversed.
This is a rare situation; it can occur only in (nearly) massless 
systems embedded within (much heavier) matter, such as spin systems
(cf. {\sc Purcell \& Pound} \cite{PurP}) or vortices in 2-dimensional 
fluids (cf. {\sc Montgomery \& Joyce} \cite{MonJ}, 
{\sc Eyinck \& Spohn} \cite{EyiS}). A massive thermal system  
couples significantly to kinetic energy. In this case, the total 
momentum $p$ is an extensive quantity, related to the velocity $v$, 
the corresponding intensive variable, by $p = M v$, where $M$ is the 
extensive total mass of the system. From \gzit{e.Sx}, we find that 
$\ol p = \Omega\partial \Delta/\partial v$, which implies that
$\Delta = \Delta|_{v=0} + \frac{\ol M}{2\Omega} v^2$. Since the mass is
positive, this expression is convex in $v$, not concave; hence $T>0$.
Thus, in a massive thermal system, the temperature must be 
positive. 

(ii) In the application, the free scaling constant $\kappa$ is usually
chosen as $\kappa=\kbar/\Omega$, where $\Omega$ is a measure of 
{\bf system size}, e.g., the total volume or total mass of the system.
In actual calculations from statistical mechanics, the integral is 
usually a function of the system size. To make the result independent
of it, one performs the so-called thermodynamic limit $\Omega\to\infty$;
thus $\Omega$ must be chosen in such a way that this limit is 
nontrivial. Extensivity in single phase global equilibrium then 
justifies treating $\Omega$ as an arbitrary positive factor. 
\end{rems}

The equation of state shows that, apart from possible singularities 
describing so-called phase transitions, the state space has the 
structure of an $(s-1)$-dimensional manifold in $\Rz ^{s}$, where 
$s$ is the number of intensive variables. 

In phenomenological thermodynamics (cf. Section \ref{s.phen}), 
one makes suitable, more or 
less heuristic assumptions on the form of the system function, while in
{\bf statistical mechanics}, one derives its form from \gzit{e.eos}
and specific choices for the quantities $H$ and $X$ within one of the
settings described in Example \ref{e3.1}. Given these choices, the 
main task is then the evaluation of the system function \gzit{e.eosa} 
since everything else can be computed from it.
Frequently, \gzit{e.eosa} can be approximately evaluated from the 
cumulant expansion \gzit{e.cumulant} and/or a mean field approximation 
\gzit{e.meanfield}.

An arbitrary Gibbs state is generally not a thermal state. However, 
we can try to approximate it by an equilibrium state.

\begin{thm} \label{t7.3}
Let $\<\cdot\>$ be a Gibbs state with entropy $S$.
Then, for arbitrary $(T,\alpha)$ satisfying $T>0$ and the equation of 
state \gzit{e.eos}, the values 
$\ol H=\<H\>$, $\ol S=\<S\>$, and $\ol X=\<X\>$ satisfy
\lbeq{e.excess}
\ol H \ge T\ol S-\alpha\cdot \ol X.
\eeq
Equality only holds if $S$ is the entropy of a thermal state with 
intensive variables $(T,\alpha)$.
\end{thm}

\bepf
The equation of state implies that $S_c:=T^{-1}(H-\alpha\cdot X)$
is the entropy of a thermal state. Now the assertion follows from 
Theorem \ref{t4.5}, since
$\<S\>\le \<S_c\>=T^{-1}(\<H\>-\alpha\cdot \<X\>)$,
with equality only if $S=S_c$.
\epf

As the theorem shows, everything of macroscopic interest is 
deducible from an explicit formula for the system function.
This explains why, in many situations, one can use 
thermodynamics very successfully as a phenomenological theory 
without having to bother about microscopic details. It suffices that
a phenomenological expression for $\Delta(T,\alpha)$ is available.
In particular, the phenomenological axioms from Section \ref{s.phen}
now follow by specializing the above to a {\bf standard system}, 
characterized by the extensive quantities
\lbeq{3-1}
H, X_0=V,\quad X_j=N_j\ (j\ne 0),
\eeq
where, as before, $V$ denotes the (positive) {\bf volume} of the 
system, and each $N_j$ denotes the (nonnegative) number of molecules 
of a fixed chemical composition (we shall call these {\bf particles of 
species} $j$). However, $H$ and the $N_j$ are now quantities from $\Ez$,
rather than thermal variables.
\lbeq{3-4}
P:=-\alpha_0 
\eeq
is called {\bf pressure}, and 
\lbeq{3-4z}
\mu _j:=\alpha_j\quad (j\not=0)
\eeq
the {\bf chemical potential} of species $j$; hence
\[
\alpha \cdot X = -PV+\mu\cdot N.
\]
Specializing the theorem, we find the phenomenological Euler equation 
\lbeq{e.Hx}
\ol{H}=T\ol{S}-P V +\mu \cdot \ol{N}.
\eeq
For reversible changes, we have the first law of thermodynamics 
\lbeq{3.1stx}
d\ol{H}=Td\ol{S}-Pd V +\mu \cdot d\ol{N}
\eeq
and the Gibbs-Duhem equation
\lbeq{e.GDx}
0=\ol{S}dT- V dP+\ol{N}\cdot d\mu.
\eeq
A comparison with Section \ref{s.phen}
shows that dropping the bars from the values 
reproduces for $T>0$, $P>0$  and $\ol S\ge 0$ the axioms of 
phenomenological thermodynamics, except for the extensivity outside 
equilibrium (which has local equilibrium as its justification).
The assumption $T>0$ 
was justified in Remark \ref{r5.2}(i), and $\ol S\ge 0$ will be 
justified in Section \ref{third}. But I have been unable to find a 
theoretical argument which shows that the pressure of a standard system 
in the above sense must always be positive. (At $T<0$, negative 
pressure is possible; see Example \ref{ex.schottky}.)
I'd appreciate getting information about this from my readers.

Apart from boundary effects, which become more and more unimportant 
as the system gets larger, the extensive quantities scale linearly with 
the volume. In the thermodynamic limit, corresponding 
to an idealized system infinitely extended in all directions, this 
becomes exact, although this can be proved rigorously only in simple 
situations, e.g., for hard sphere model systems
({\sc Yang \& Lee} \cite{YanL}) or spin systems 
({\sc Griffiths} \cite{Gri}).
A thorough treatment of the thermodynamic limit (e.g., {\sc Thirring} 
\cite{Thi}, or, in the framework of large deviation theory, 
{\sc Ellis} \cite{Ell}) in general needs considerably more algebraic 
and analytic machinery, e.g., the need to work in place of thermal 
states with more abstract KMS-states (which are limits of sequences of 
thermal states still satisfying a KMS condition \gzit{e.KMS}).
Moreover, proving the existence of the limit requires detailed 
properties of the concrete microscopic description of the system.

For very small systems, typically atomic clusters or molecules, 
$N$ is fixed and a {\bf canonical ensemble} without the $\mu \cdot N$ 
term is more appropriate. For the thermodynamics of small systems 
(see, e.g., ({\sc Bustamente} et al. \cite{BusLR}, 
{\sc Gross} \cite{Gro}, {\sc Kratky} \cite{Kra})
such as a single cluster of atoms, $V$ is still taken as a fixed 
reference volume, but now changes in the physical volume (adsorption 
or dissociation at the surface) are not represented in the system, 
hence need not respect the thermodynamic laws. For large 
surfaces (e.g., adsorption studies in chromatography \cite{KarSH,Mas}), 
a thermal description is achievable by including additional variables 
(surface area and surface tension) to account for the boundary effects;
but clearly, surface terms scale differently with system size 
than bulk terms.

Thus, whenever the thermal description is valid, computations can be 
done in a fixed reference volume $V_0$ which we take as system size
$\Omega$, and the true, variable volume $V$ 
can always be represented in the Euclidean *-algebra as a real number, 
so that in particular $\ol V=V$. Then \gzit{e.eosa} implies that, for 
the reference volume,
\[
\Delta(T,\alpha) = \Omega^{-1}\kbar T
\log (e^{-\beta P\Omega})\sint e^{-\beta (H-\mu\cdot N)}),
\]
hence 
\lbeq{e.eosb}
\Delta(T,\alpha)= \Omega^{-1}\kbar T(\log Z(T,\mu) - P\Omega)
=P(T,\mu)-P,
\eeq
where
\lbeq{3-10}
Z(T,\mu):=\sint e^{-\beta (H-\mu \cdot N)}
\eeq
is the so-called {\bf grand canonical partition function} of the system
and 
\lbeq{3-9a}
P(T,\mu):=\Omega^{-1}\kbar T\log Z(T,\mu). 
\eeq
With our convention of considering a fixed reference volume and treating
the true volume $V$ as a scale factor (otherwise a thermodynamic limit 
would be needed), this expression is independent of $V$, since it 
relates intensive variables unaffected by scaling. 
The equation of state \gzit{e.eos} therefore takes the form
\lbeq{e.eos2}
P= P(T,\mu).
\eeq
Quantitative expressions for the equation of state can often be 
computed from \gzit{3-10}--\gzit{3-9a} using the cumulant expansion 
\gzit{e.cumulant} and/or a mean field approximation \gzit{e.meanfield}.
Note that these relations imply
\[
e^{-\beta P(T,\mu)V} = \sint e^{-\beta(H-\mu\cdot N)}.
\]
Traditionally (see, e.g., {\sc Gibbs} \cite{Gib}, 
{\sc Huang} \cite{Hua}, {\sc Reichl} \cite{Rei}), 
the thermal state corresponding to \gzit{e.eosb}--\gzit{3-9a}
is called a {\bf grand canonical ensemble}, and the following results 
are taken as the basis for microscopic calculations from 
statistical mechanics.

\begin{thm} \label{3.3.}
For a standard system in global equilibrium, values of an
arbitrary quantity $g$ can be calculated from \gzit{3-10} and
\lbeq{3-11}
\< g\> =Z(T,\mu)^{-1}\sint e^{-\beta (H-\mu \cdot N)}g.
\eeq
The values of the extensive quantities are given in terms of 
the equation of state \gzit{3-9a} by
\lbeq{3-19}
\ol{S}= V \frac{\partial P}{\partial T}(T,\mu ),~~~
\ol{N}_j= V \frac{\partial P}{\partial \mu _j}(T,\mu )
\eeq
and the phenomenological Euler equation \gzit{e.Hx}.
\end{thm}

\bepf 
Equation \gzit{3-9a} implies that
\[ 
\begin{array}{lll}
\<g\> &=\sint e^{-S/\kbar }g=\sint e^{-\beta (H+PV-\mu \cdot N)}g \\ 
&=e^{-\beta PV}\sint e^{-\beta (H-\mu \cdot N)}g=
Z(T,V,\mu )^{-1} \sint e^{-\beta (H-\mu \cdot N)}g,
\end{array} 
\]
giving \gzit{3-11}. The formulas in \gzit{3-19} follow from 
\gzit{e.Sx} and \gzit{e.eosb}. 
\epf

No thermodynamic limit was needed to derive the above results. 
Thus, everything holds -- though with large limit resolutions in 
measurements -- even for single small systems 
({\sc Bustamente} et al. \cite{BusLR}, {\sc Gross} \cite{Gro}, 
{\sc Kratky} \cite{Kra}).

\begin{expl}\label{ex.schottky}
We consider the two level system from Example \ref{ex.canonical},
using $\Omega=1$ as system size. From \gzit{3-10} and  \gzit{3-9a},
we find $Z(T,\mu)=1+e^{-E/\kbar T}$, hence
\[
P(T,\mu)=\kbar T\log(1+e^{-E/\kbar T})
=\kbar T\log(e^{E/\kbar T}+1)-E.
\] 
From \gzit{3-11}, we find 
\[
\ol H = \frac{Ee^{-E/\kbar T}}{1+e^{-E/\kbar T}}
=\frac{E}{e^{E/\kbar T}+1},~~~\kbar T = \frac{E}{\log(E/\ol H -1)}.
\]
(This implies that a two-level system has negative temperature and 
negative pressure if $\ol H>E/2$.)
The {\bf heat capacity} $C:=d\ol H/dT$ takes the form
\[
C=\frac{E^2}{\kbar T^2}\frac{e^{E/\kbar T}}{(e^{E/\kbar T}+1)^2}.
\]
It exhibits a pronounced maximum, the so-called {\bf Schottky bump}
(cf. {\sc Callen} \cite{Cal}), from which $E$ can be determined.
In view of \gzit{e.2levelapprox} below, this allows the experimental
estimation of the spectral gap of a quantum system.
The phenomenon persists to some extent for
multilevel systems; see {\sc Civitarese} et al. \cite{CivHH}.
\end{expl}

\section{Description levels}\label{s.detail}

So far, we have assumed a fixed selection of extensive quantities
defining the thermal model. However, as indicated at the end
of Section \ref{s.phen}, observable differences from the conclusions
derived from a thermal model imply that one or more conjugate 
pairs of thermal variables are missing in the model.
In this section, we discuss in more detail the relation between 
different thermal models constructed on the basis of the same 
Euclidean *-algebra by selecting different lists of extensive 
quantities.

Our first observation is the flexibility of the thermal setting.
While the zeroth law may look very restrictive at first sight, by 
choosing a large enough family of extensive quantities the entropy of 
an {\em arbitrary}\/ Gibbs state can be approximated arbitrarily well
by a linear combination of these quantities.

The zeroth law thus simply appears as an embodiment of {\sc Ockham}'s 
razor\footnote{\label{f.ockham}
frustra fit per plura quod potest fieri per pauciora
}\ 
 \cite{Ock}, freely paraphrased in modern form: that we should opt for 
the most economic model explaining a phenomenon -- by restricting 
attention to the relevant extensive quantities only.
At each time $t$, there is -- except in degenerate cases --
a {\em single} Gibbs state, with entropy $S(t)$, say, which best 
describes the system under consideration. Assuming the description
by the Gibbs state as fundamental, its value is the objective,
true value of the entropy, relative only to the algebra of quantities 
chosen to model the system. A description of the state 
in terms of a thermal system is therefore adequate if (and, under
an observability qualification to be discussed below, only if), 
for all relevant times $t$, the entropy $S(t)$ can be adequately 
approximated by a linear combination of the extensive quantities 
available at the chosen level of description. 

The set of extensive variables depends on the application 
and on the desired accuracy of the model; it must be chosen in such 
a way that knowing the measured values of the extensive variables 
determines (to the accuracy specified) the complete behavior of the
thermal system. Thus, the choice of extensive 
variables is (to the accuracy specified)
completely determined by the level of accuracy with which 
the thermal description should fit the system's behavior. 
This forces everything else: 
The theory must describe the freedom available to
characterize a particular thermal system with this set of
extensive variables, and it must describe how the numerical values of
interest can be computed for each state of each thermal system.

In contrast to the information theoretic approach where the choice of 
extensive quantities is considered to be the subjective matter of which 
observables an observer happens to have knowledge of, the only
subjective aspect of our setting is the choice of the 
resolution of modeling. This fixes the amount of approximation 
tolerable in the ansatz, and hence the necessary list of extensive 
quantities. (Clearly, physics cannot be done without approximation,
and the choice of a resolution is unavoidable.
To remove even this trace of subjectivity, inherent in
any approximation of anything, the entropy would have to be 
represented without any approximation, which would require to use 
as the algebra of quantities the still unknown theory of everything,
and to demand that the extensive quantities exhaust this algebra.)

\begin{table}[htb]
\caption{Typical conjugate pairs of thermal variables
and their contribution to the Euler equation. 
The signs are fixed by tradition. (in the gravitational term,
$m$ is the vector with components $m_j$, the mass of a particle of 
species $j$, $g$ the acceleration of gravity, and $h$ the height.)
}
\label{3.t.}

\begin{center}
\begin{tabular}{|l|l|l|}
\hline
extensive $X_j$ & intensive $\alpha_j$ 
& contribution $\alpha_jX_j$ \\
\hline
\hline
entropy $S$ & temperature $T$ 
& thermal, $TS$ \\
\hline
particle number $N_j$ & chemical potential $\mu _j$ 
& chemical, $\mu _jN_j$\\
conformation tensor $C$ & relaxation force $R$
& conformational $\sum R_{jk} C^{jk}$\\
\hline
strain $\eps^{jk}$ & stress $\sigma_{jk}$
& elastic, $\sum \sigma_{jk}\eps^{jk}$\\
volume $V$ & pressure $-P$ 
& mechanical, $-PV$ \\
surface $A_S$ & surface tension $\gamma $ 
& mechanical, $\gamma A_S$ \\
lenght $L$ & tension $J$ 
& mechanical, $JL$ \\
displacement $q$ & force $-F$
& mechanical, $-F\cdot q$\\
momentum $p$ & velocity $v$
& kinetic, $v\cdot p$ \\
angular momentum $J$ & angular velocity $\Omega$
& rotational, $\Omega \cdot J$\\
\hline
charge $Q$ & electric potential $\Phi$
& electrical, $\Phi Q$\\
polarization $P$ & electric field strength $E$ 
& electrical, $E\cdot P$\\
magnetization $M$ & magnetic field strength $B$ 
& magnetical, $B\cdot M$ \\
e/m field $F$ & e/m field strength $-F^s$ 
& electromagnetic, $-\sum F^s_{\mu\nu}F^{\mu\nu}$\\
\hline
mass $M=m\cdot N$ & gravitational potential $gh$
& gravitational, $ghM$ \\
energy-momentum $U$ & metric $g$ 
& gravitational, $\sum g_{\mu\nu}U^{\mu\nu}$ \\
\hline
\end{tabular}
\end{center}
\end{table}

In general, which quantities need 
to be considered depends on the resolution with which the system is 
to be modeled -- the higher the resolution the larger the family of 
extensive quantities. 
Thus -- whether we describe bulk matter, surface effects, 
impurities, fatigue, decay, chemical reactions, or transition states, 
-- the thermal setting remains the same since it is a universal 
approximation scheme, while the number of degrees of freedom 
increases with increasingly detailed models. 

In phenomenological thermodynamics, the relevant extensive 
quantities are precisely those variables that are
observed to make a difference in modeling the phenomenon of interest.
Table \ref{3.t.} gives typical extensive variables, their intensive 
conjugate variables, and their contribution to the Euler 
equation\footnote{\label{f.euler} 
The Euler equation, which contains the energy 
contributions specified in the table, looks like an  
energy balance. But since $S$ is undefined, this formal balance has no 
contents apart from defining the entropy $S$ in terms of the energy 
and other contributions. 
The energy balance is rather given by the first law discussed later, 
and is about {\em changes} in energy. Conservative work contributions 
are exact differentials. For example, the mechanical force 
$F=-dV(q)/dq$ translates into the term 
$-F\cdot dq=dV(q)$ of the first law, corresponding to the
term $-F\cdot q$ in the Euler equation.
The change of the kinetic energy $E_\kin=mv^2/2$ contribution of 
linear motion with velocity $v$ and momentum $p=mv$ is 
$dE_\kin=d(mv^2/2)= mv \cdot dv = v \cdot dp$, which is exactly what 
one gets from the $v \cdot p$ contribution in the Euler 
equation. Since $v \cdot p = mv^2$ is larger than the kinetic
energy, this shows that motion implies a contribution to the
entropy of $(E_\kin-v \cdot p)/T=-mv^2/2T$.
A similar argument applies to the angular motion of a rigid 
body in its rest frame, providing the term involving angular velocity 
and angular momentum.
}. 
Some of the extensive variables and their intensive conjugates are 
vectors or (in elasticity theory, the theory of complex fluids, and in 
the relativistic case) tensors; cf. {\sc Balian} \cite{Bal} for the 
electromagnetic field and {\sc Beris \& Edwards} \cite{BerE}, 
{\sc \"Ottinger} \cite{Oet} for complex fluids.

\bigskip
To analyze the relation between two different thermal description 
levels, we compare a coarse system and a more detailed system 
quantitatively, taking for simplicity the temperature constant, 
so that the $T$-dependence can be suppressed in the formulas,
and states are completely determined by $\alpha$.

The fine system will be written as before, the variables and 
quantities associated with the coarser system get an additional 
index $c$. In order to be able to compare the two systems, we assume 
that one is a refinement of the other, so that the extensive quantities 
of the coarse system are $X_c=CX$, with a fixed matrix $C$ with 
linearly independent rows, whose components tell how the components of 
$X_c$ are built from those of $X$. The entropy of the coarse system
is then given by
\[
S_c=T^{-1}(H-\alpha_c\cdot X_c)
=T^{-1}(H-\alpha_c\cdot CX)
=T^{-1}(H-\alpha^*\cdot X),
\]
where 
\lbeq{e.star}
\alpha^*=C^T\alpha_c.
\eeq
Thus, the thermal states of the coarse model are simply the states of 
the detailed model for which the intensive parameter vector $\alpha$ 
is of the form $\alpha=C^T\alpha_c$ for some $\alpha_c$. Thus the
coarse state space can simply be viewed as a lower-dimensional subspace 
of the detailed state space. Therefore one expects the coarse 
description to be adequate precisely when the detailed state is close 
to the coarse state space, with an accuracy determined by the 
desired fidelity of the coarse model. 
Since the relative entropy \gzit{4-5},
\lbeq{e.relent}
\< S_c-S\>= \<T^{-1}(H-\alpha_c\cdot CX)-T^{-1}(H-\alpha\cdot X)\>
=\<T^{-1}(\alpha-C^*\alpha_c)\cdot X\>,
\eeq
measures the amount of 
information in the detailed state which cannot be explained by
the coarse state, it is sensible to associate to an arbitrary detailed
state $\alpha$ the coarse state $\alpha_c$ determine by minimizing
\gzit{e.relent}. If $\alpha^*=C^T\alpha_c\approx \alpha$ then 
\[
S_c=T^{-1}(H-\alpha^*\cdot X)\approx 
T^{-1}(H-\alpha^*\cdot X)=S,
\]
and the coarse description is adequate. 
If $\alpha^*\not\approx \alpha$, there is no a priori reason to trust
the coarse model, and we have to investigate to which extent its 
predictions will significantly differ from those of the detailed model.
One expects the differences to be significant; however, 
in practice, there are difficulties if there are limits on our
ability to prepare particular detailed states. The reason is that
the entropy and chemical potentials can be prepared and measured
only by comparison with states of sufficiently known states.
A first sign of this is the gauge freedom in ideal gases discussed
in Example \ref{ex.ideal}, which implies that different models of the 
same situation may have nontrivial differences in Hamilton energy, 
entropy, and chemical potential. This ambiguity persists in more 
perplexing situations:

\begin{expl} \label{ex.gibbs}
{\bf (The Gibbs paradox)}\\
Suppose that we have an ideal gas of two 
species $j=1,2$ of particles which are experimentally indistinguishable.
Suppose that in the samples available for experiments, the two species 
are mixed in significantly varying proportions $N_1:N_2=q_1:q_2$ which,
by assumption, have no effect on the observable properties; in 
particular, their values are unknown but varying.
The detailed model treats them as distinct, the coarse model as 
identical. 
Reverting to the barless notation of Section \ref{s.phen}, we have
\[
X=\pmatrix{V\cr N_1 \cr N_2},~~~\alpha = \pmatrix{-P\cr \mu_1\cr \mu_2},
\]
and, assuming $C=\pmatrix{1 & 0 & 0\cr 0 & c_1 & c_2}$ for suitable
$c_1,c_2>0$,
\[
X_c=\pmatrix{V\cr N_c}=\pmatrix{V\cr c_1N_1+c_2N_2},~~~
\alpha_c= \pmatrix{-P \cr \mu_c}.
\]
From the known proportions, we find
\[
N_j=x_jN_x,~~~ x_j=\frac{q_j}{c_1q_1+c_2q_2}.
\]
The mixture behaves like an ideal gas of a single species, hence 
\[
PV=\kbar T N_c,~~~H=h_c(T)N_c,~~~
\mu_c=\kbar T \log\frac{\kbar TN_c}{V\pi_c}.
\]
Now $N_c=(\kbar T)^{-1}PV=\sum N_j=\sum x_j N_c$ implies that
$x_1+x_2=1$. Because of indistinguishability, this must hold for
any choice of $q_1,q_2\ge 0$; for the two choices $q_1=0$ and $q_2=0$,
we get $c_1=c_2=1$, hence $N_c=\sum N_j$, and the $x_j$ are mole 
fractions. Similarly, if we use for all species $j$ the same 
normalization for fixing the gauge freedom, the relation 
$h_c(T)N_c=H=\sum h_j(T)N_j=\sum h_j(T)x_jN_c$ implies for varying 
mole fractions that $h_j(T)=h_c(T)$ for $j=1,2$. From this, we get 
$\pi_j(T)=\pi_c(T)$ for $j=1,2$. Thus
\[
H-H_c = \sum h_j(T)N_j - h_c(T)N_c = 0,
\]
\[
\mu_j-\mu_c =\kbar T \log\frac{\kbar TN_j}{V\pi_j}-
\kbar T \log\frac{\kbar TN_c}{V\pi_c}
=\kbar T \log x_j,
\]
the Gibbs energy satisfies
\[
G-G_c=\sum \mu_jN_j-\mu_cN_c = \sum (\mu_j-\mu_c)N_j
=\kbar T N_c \sum x_j\log x_j,
\]
and the entropy satisfies 
\[
\bary{lll}
S-S_c&=&T^{-1}(H-PV+G) - T^{-1}(H_c-PV+G_c) \\
&=& T^{-1}(G-G_c) =\kbar N_c \sum x_j\log x_j.
\eary
\]
This term is called the {\bf entropy of mixing}.
Its occurence is referred to as Gibbs paradox
(cf. {\sc Jaynes} \cite{Jay.G}, {\sc Tseng \& Caticha} \cite{TseC}, 
{\sc Allahverdyan \& Nieuwenhuizen} \cite{AllN},
{\sc Uffink} \cite[Section 5.2]{Uff}).
It seems to say that there are two different entropies, depending on 
how we choose to model the situation. 
For fixed mole fractions, the paradox can be 
resolved upon noticing that the fine and the coarse description 
differ only by a choice of gauge; the gauge is unobservable anyway, 
and the entropy is determined only when the gauge is fixed.

However, if the mole fractions vary, the fine and the coarse 
description differ significantly. If the detailed model is correct, 
the coarse model gives a wrong description for the entropy and the 
chemical potentials. However, this difference is observable only if we
know processes which affect the different species differently, 
such as a difference in mass, which allows a mechanical separation, 
in molecular size or shape, which allows their separation by a 
semipermeable membrane, in spin, which allows a magnetic separation, 
or in scattering properties of the particles, which may allow a
chemical or radiation-based differentiation.
In each of these cases, the particles become distinguishable, 
and the coarse description becomes inadequate. 

If we cannot separate the species to some extent, we cannot prepare
equilibrium ensembles at fixed mole fraction. But this would be 
necessary to calibrate the chemical potentials, since fixed chemical
potentials can be prepared only through chemical contact with
substances with known chemical potentials, and the latter must be
computed from mole fractions.
\end{expl}

Generalizing from the example, we conclude that even when both a 
coarse model and a 
more detailed model are faithful to all experimental information 
possible at a given description level, there is no guarantee that 
they agree in the values of all thermal variables of the
coarse model. In the language of control theory (see, e.g., 
{\sc Ljung} \cite{Lju}), agreement is guaranteed only when all  
parameters of the more detailed models are observable. 

On the other hand, all observable state functions of the detailed 
system which depend only on the coarse state must and will have the 
same value within the experimental accuracy if both models are adequate
descriptions of the situation. Thus, while the values of some variables 
need not be experimentally determinable, the validity of a model is an 
objective property. Therefore, preferences for one or the other of two 
valid models can only be based on other criteria. The criterion usually
employed in this case is Ockham's 
razor$^{\mbox{\protect\ref{f.ockham}}}$, although
there may be differences of opinion on what counts as the most
economic model. In particular, a fundamental description of macroscopic 
matter by means of quantum mechanics is hopelessly overspecified in 
terms of the number of degrees of freedom needed for comparison with 
experiment, most of which are in principle unobservable by equipment 
made of ordinary matter. But it is often the most economical model in 
terms of description length (though extracting the relevant 
information from it may be difficult).
Thus, different people may well make different rational choices,
or employ several models simultaneously.

The objectivity of a model description implies that, as soon as a 
discrepancy with experiment is reliably found, the model must be
replaced by a more detailed (or altogether different) model. 
This is indeed what happened with the textbook example of the
Gibbs paradox situation, ortho and para hydrogen, cf. 
{\sc Bonhoeffer \& Harteck} \cite{BonH}, {\sc Farkas} \cite{Far}.
Hydrogen seemed at first to be a single substance, but then 
thermodynamic data forced a refined description.

Similarly, in spin echo experiments (see, e.g., 
{\sc Hahn} \cite{Hah,Hah2}, {\sc Rothstein} \cite{Rot},
{\sc Ridderbos \& Redhead} \cite{RidR}), the specially 
prepared system appears to be in equilibrium but, according to 
Callen's empirical definition$^{\mbox{\protect\ref{f.Callen}}}$ 
it is not -- 
the surprising future behavior (for someone not knowing the special 
preparation) shows that some correlation variables were 
neglected that are needed for a correct description. 
Indeed, everywhere in science, we strive for explaining surprising 
behavior by looking for the missing variables needed to describe the 
system correctly!

{\sc Grad} \cite{Grad} expresses this as ''the adoption of a new 
entropy is forced by the discovery of new information''.
More precisely, the adoption of a new {\em model} is forced, since the 
old model is simply wrong under the new conditions and remains valid 
only under some restrictions. 
Thus this is not a property of entropy alone, but of all concepts in 
models of reality relating to effects not observable (in the sense
of control theory discussed above).

Observability issues aside, the coarser description usually has a more 
limited range of applicability; with the qualification discussed in 
the example, it is generally restricted to those systems whose 
detailed intensive variable vector $\alpha$ is close to 
the subspace of vectors of the form $C^T\alpha$ reproducible in the 
coarse model.
Finding the right family of thermal variables is therefore a 
matter of discovery, not of subjective choice. This is further 
discussed in Section \ref{s.model}.

\section{The first law: Energy balance} \label{first}

We now discuss relations between 
changes of the values of extensive or intensive variables,
as expressed by the first law of thermodynamics. To derive the first 
law in full generality, we use the concept of reversible 
transformations introduced in Section \ref{s.phen}. Corresponding
to such a transformation, there is a family of thermal states 
$\<\cdot\>_\lambda$ defined by
\[
\<f\>_\lambda = \sint e^{-\beta(\lambda)(H-\alpha(\lambda)\cdot X)}f,~~~
\beta(\lambda)=\frac{1}{\kbar T(\lambda)}.
\]
{\bf Important:} In case of local or microlocal equilibrium, 
where the thermal system carries a dynamics, it is important
to note that reversible transformations are ficticious transformations
which have nothing to do with how the system changes with time,
or whether a process is reversible in the dynamical sense that both
the process and the reverse process can be realized dynamically.
The time shift is generally {\em not} a reversible transformation.

We use differentials corresponding to reversible transformations;
writing $f=S/\kbar$, we can delete the index $f$ from the formulas
in Section \ref{s.gibbs}. In particular, we write the Kubo 
inner product \gzit{e.kubo} as
\lbeq{e.kuboS}
\<g;h\> := \<g;h\>_{S/\kbar}.
\eeq

\begin{prop}
The value $\ol g(T,\alpha):=\<g(T,\alpha)\>$ of every 
(possibly $T$- and $\alpha$-dependent) quantity 
$g(T,\alpha)$ is a state variable satisfying the  {\bf differentiation 
formula}
\lbeq{e.diffS}
d\<g\>=\<dg\>-\<g-\ol g;dS\>/\kbar.
\eeq
\end{prop}
\bepf
That $\ol g$ is a state variable is an immediate consequence of the 
zeroth law \gzit{3-2} since the entropy depends on $T$ and $\alpha$ 
only. The differentiation formula follows from \gzit{e.diff} and
\gzit{e.kuboS}.
\epf

\begin{thm} \label{3.6.}
For reversible changes, we have the {\bf first law of thermodynamics}
\lbeq{3.1st}
d\ol{H}=Td\ol{S}+\alpha\cdot d\ol{X}
\eeq
and the {\bf Gibbs-Duhem equation}
\lbeq{e.GD}
0=\ol{S}dT+\ol{X}\cdot d\alpha.
\eeq

\end{thm}

\bepf 
Differentiating the equation of state \gzit{e.eos}, using the chain 
rule \gzit{3-12}, and simplifying using \gzit{e.Sx} gives the 
Gibbs-Duhem equation (\ref{e.GD}). If we differentiate the 
phenomenological Euler equation \gzit{e.H}, we obtain
\[
d\ol{H}=Td\ol{S}+\ol{S}dT+\alpha\cdot d\ol{X}+
\ol{X}\cdot d\alpha,
\]
and using (\ref{e.GD}), this simplifies to the first law of
thermodynamics.
\epf

Because of the form of the energy terms in the first law (\ref{3.1st}), 
one often uses the analogy to mechanics and calls the intensive 
variables {\bf generalized forces}, and differentials of extensive 
variables {\bf generalized displacements}.

For the Gibbs-Duhem equation, we give a second proof which provides 
additional insight. Since $H$ and $X$ are fixed
quantities for a given system, they do not change under reversible
transformations; therefore 
\[
dH=0,~~~dX = 0.
\]
Differentiating the Euler equation \gzit{e.euler}, therefore gives 
the relation
\lbeq{3-16}
0=TdS+SdT+X\cdot d\alpha.
\eeq
On the other hand, $S$ depends explicitly on $T$ and $\alpha$, 
and by Corollary \ref{app2.},
\lbeq{3-17}
\< dS\> =\int e^{-S/\kbar }dS=
\kbar d \left(\int e^{-S/\kbar }\right)=
\kbar d1=0,
\eeq
taking values in (\ref{3-16}) implies again the Gibbs-Duhem
equation. Equation \gzit{3-16} can also be used to get
information about limit resolutions.

\begin{thm} \label{7.1.}~\\
(i) Let $g$ be a quantity depending continuously differentiable on the 
intensive variables $T$ and $\alpha$. Then
\lbeq{7-4}
\< g-\ol{g};S-\ol{S}\> =
\kbar T\Big(\frac {\partial \ol{g}} {\partial T}-
\Big\<\frac{\partial g}{\partial T}\Big\>\Big), 
\eeq
\lbeq{7-5}
\<g-\ol{g};X_j-\ol{X}_j\> =
\kbar T\Big(\frac {\partial \ol{g}} {\partial \alpha_j}-
\Big\<\frac{\partial g}{\partial\alpha_j}\Big\> \Big),
\eeq
(ii) If the extensive variables $H$ and $X_j$ ($j\in J$) are pairwise 
commuting then
\lbeq{7-7}
\< (S-\ol{S})^2\> =
\kbar T\ \frac {\partial\ol{S}} {\partial T},
\eeq
\lbeq{7-8}
\< (X_j-\ol{X}_j)
(S-\ol{S})\> =\kbar T
\ \frac {\partial\ol{X}_j} {\partial T}\quad \quad (j\in J),
\eeq
\lbeq{7-9}
\< (X_j-\ol{X}_j)
(X_k-\ol{X}_k)\> =\kbar T
\ \frac {\partial\ol{X}_j} {\partial\alpha_k}\quad \quad (j,k\in J),
\eeq
\lbeq{e.limitSX}
\res(S)
=\sqrt{\frac{\kbar T}{\ol S^2}\frac{\partial \ol S}{\partial T}},~~~
\res(X_j)
=\sqrt{\frac{\kbar T}{\ol X_j^2}
\frac{\partial \ol X_j}{\partial \alpha_j}},
\eeq
\lbeq{e.limitH}
\res(H)
=\sqrt{\frac{\kbar T}{\ol H^2}
\Big(T\frac{\partial \ol H}{\partial T}+
\alpha\cdot\frac{\partial \ol H}{\partial \alpha}\Big)}.
\eeq
\end{thm}

\bepf 
Multiplying the differentiation formula \gzit{e.diffS} by $\kbar T$ 
and using \gzit{3-16}, we find, for arbitrary reversible 
transformations,
\[
\kbar T(d\<g\>-\<dg\>)=\<g-\ol g;S\>dT + \<g-\ol g;X\>\cdot d\alpha.
\]
Dividing by $d\lambda$ and choosing $\lambda =T$ and 
$\lambda =\alpha_j$, respectively, gives
\[
\< g-\ol{g};S\> =
\kbar T\Big(\frac {\partial \ol{g}} {\partial T}-
\Big\<\frac{\partial g}{\partial T}\Big\>\Big),~~~
\<g-\ol{g};X_j\> =
\kbar T\Big(\frac {\partial \ol{g}} {\partial \alpha_j}-
\Big\<\frac{\partial g}{\partial\alpha_j}\Big\> \Big).
\]
(i) follows upon noting that $\<g-\ol g;h-\ol h\>= \<g-\ol g;h\>$ since
by \gzit{e.kubo1},
\[
\<g-\ol g;\ol h\>=\<g-\ol g\>\ol h=(\<g\>-\ol g)=0.
\]
If the extensive variables $H$ and $X_j$ ($j\in J$) are pairwise 
commuting then we can use \gzit{e.kubo0} to eliminate the Kubo inner
product, and by choosing $g$ as $S$ and $X_j$, respectively, we find 
\gzit{7-7}--\gzit{7-9}. The limit resolutions \gzit{e.limitSX} now
follow from \gzit{e.res} and the observation that $\<(g-\ol g)^2\>
=\<(g-\ol g)g\>-\<g-\ol g\>\ol g = \<(g-\ol g)g\> =\<g^2\>-\ol g^2$.
The limit resolution \gzit{e.limitH} follows similarly from
\[
\bary{lll}
\res(H)^2&=&\<H-\ol H;H-\ol H\> 
= T\<H-\ol H;S-\ol S\>+\alpha\cdot\< H-\ol H;X-\ol X\>\\
&=&\D\kbar T \Big(T\frac{\partial \ol H}{\partial T}+
\alpha\cdot\frac{\partial \ol H}{\partial \alpha_j}\Big).
\eary
\] 
\epf

Note that higher order central moments can be obtained in the same 
way, substituting more complicated expressions for $f$ and using the
formulas for the lower order moments to evaluate the right hand side
of (\ref{7-4}) and (\ref{7-5}). 

The extensive variables scale linearly
with the system size $\Omega$ of the system. Hence, the limit 
resolution of the extensive quantities is $O(\sqrt{\kbar/\Omega})$ in 
regions of the state space where the
extensive variables depend smoothly on the intensive variables. 
Since $\kbar$ is very small, they are negligible unless the system 
considered is very tiny. Thus, macroscopic thermal variables 
can generally be obtained with fairly high precision.
An exception is close to critical points 
where the extensive variables are not differentiable, 
and their derivatives can therefore become huge. 
In particular, in the thermodynamic limit $\Omega\to\infty$, 
uncertainties are absent except close to a critical point, where
they lead to critical opacity.

\begin{cor} \label{7.4.}
For a standard thermal system, 
\lbeq{e.limitSN}
\res(S)
=\sqrt{\frac{\kbar T}{\ol S^2}\frac{\partial \ol S}{\partial T}},~~~
\res(N_j)
=\sqrt{\frac{\kbar T}{\ol N_j^2}
\frac{\partial \ol N_j}{\partial \mu_j}},
\eeq
\lbeq{e.limitH2}
\res(H)
=\sqrt{\frac{\kbar T}{\ol H^2}
\Big(T\frac{\partial \ol H}{\partial T}
+P\frac{\partial \ol H}{\partial P}
+\mu\cdot\frac{\partial \ol H}{\partial \mu}\Big)}.
\eeq
\end{cor}

\bepf
Use (\ref{7-7}) and (\ref{7-9}).
\epf

Note that $\res(V)=0$ since we regarded $V$ as a number.

\section{The second law: Extremal principles} \label{second}

The extremal principles of the second law of thermodynamics
assert that in a nonthermal state, some energy expression depending 
on one of a number of standard  boundary conditions
is strictly larger than that of related thermal states.
The associated thermodynamic potentials can
be used in place of the system function to calculate all 
thermal variables given half of them.
Thus, like the system function, thermodynamic potentials give a 
complete summary of the equilibrium properties of homogeneous materials.
We only discuss the {\bf Hamilton potential}
\[
U(\ol S,\ol X)
:=\max_{T,\alpha}\,
\{T\ol S+\alpha\cdot \ol X\mid \Delta(T,\alpha)=0, T>0\}
\]
and the {\bf Helmholtz potential}
\[
 A(T,\ol X)
:=\max_\alpha\,\{\alpha\cdot \ol X\mid \Delta(T,\alpha)=0\};
\]
other potentials can be handled in a similar way.

\begin{thm} \label{t.var}
{\bf (Second law of thermodynamics)}\\
(i) In an arbitrary state, 
\[
\ol H \ge U(\ol S,\ol X),
\]
with equality iff the state is a thermal state of positive 
temperature. The remaining thermal variables are then given by
\lbeq{e.entint1}
T = \frac{\partial U}{\partial \ol S}(\ol S,\ol X),~~~
\alpha = \frac{\partial U}{\partial \ol X}(\ol S,\ol X),
\eeq
\lbeq{e.entint2}
U=\ol H = U(\ol S,\ol X).
\eeq
In particular, a thermal state of positive temperature is uniquely 
determined by the values of $\ol S$ and $\ol X$.

(ii) Let $T>0$. Then, in an arbitrary state, 
\[
\ol H-T\ol S \ge A(T,\ol X),
\]
with equality iff the state is a thermal state of temperature $T$.
The remaining thermal variables are then given by
\lbeq{e.enthelm1}
\ol S=-\frac{\partial A}{\partial T}(T,\ol X),~~~
\alpha=\frac{\partial A}{\partial \ol X}(T,\ol X),
\eeq
\lbeq{e.enthelm2}
\ol H=T \ol S + \alpha \cdot \ol X = A(T,\ol X)+T\ol S.
\eeq
In particular, a thermal state of positive temperature is uniquely 
determined by the values of $T$ and $\ol X$.
\end{thm}

\bepf 
This is proved in the same way as Theorem \ref{t.extstd};
thus we give no details.
\epf

The additivity of extensive quantities is again reflected in 
corresponding properties of the thermodynamic potentials:

\begin{thm}~\\
(i) The function $U(\ol S,\ol X)$ is a convex function 
of its arguments which is positive homogeneous of degree 1, i.e., 
for real $\lambda,\lambda^1,\lambda^2\ge 0$,
\lbeq{e.homU}
U(\lambda \ol S,\lambda \ol X)=\lambda U( \ol S,\ol  X),
\eeq
\lbeq{e.convU}
U(\lambda^1 \ol S^1+\lambda^2\ol S^2,\lambda^1 \ol X^1+\lambda^2\ol X^2)
\le \lambda^1 U(S^1,X^1)+\lambda^2 U(S^2,X^2).
\eeq

(ii) The function $A(T, \ol X)$ is a convex function 
of $X$ which is positive homogeneous of degree 1, i.e., 
for real $\lambda,\lambda^1,\lambda^2\ge 0$,
\lbeq{e.homA}
A(T,\lambda \ol X)=\lambda A(T,\ol X),
\eeq
\lbeq{e.convA}
A(T,\lambda^1 \ol X^1+\lambda^2\ol X^2)
\le \lambda^1 A(T,X^1)+\lambda^2 A(T,X^2).
\eeq
\end{thm}

\bepf  
This is proved in the same way as Theorem \ref{t.ext};
thus we give no details.
\epf

The extremal principles imply energy dissipation properties for 
time-dependent states. Since the present kinematical setting does not 
have a proper dynamical framework, it is only possible to outline the 
implications without going much into details.

\begin{thm} \label{4.4.}~\\
(i) For any time-dependent system for which $S$ and $X$ remain
constant and which converges to a thermal state with positive 
temperature, the Hamilton energy $\<H\>$ attains its global minimum in 
the limit $t\to\infty$.

(ii) For any time-dependent system maintained at fixed temperature 
$T>0$, for which $X$ remains constant and which converges to a thermal 
state, the Helmholtz energy $\<H-TS\>$ attains its global minimum 
in the limit $t\to\infty$.
\end{thm}

\bepf
This follows directly from Theorem \ref{t.var}.
\epf

This result is the shadow of a more general, dynamical observation
(that, of course, cannot be proved from kinematic assumptions alone 
but would require a dynamical theory).
Indeed, it is a universally valid empirical fact that in all natural
time-dependent processes, energy is lost or dissipated, i.e., 
becomes macroscopically unavailable, unless compensated by energy
provided by the environment. Details go beyond the present framework,
which adopts a strictly kinematic setting.

\section{The third law: Quantization} \label{third}

The third law of thermodynamics asserts that the value of 
the entropy is always nonnegative. 
But it cannot be deduced from our axioms without making
a further assumption, as a simple example demonstrates.

\begin{expl} \label{5.1.} 

The algebra $\Ez=\Cz^m$ with pointwise operations is a
Euclidean *-algebra for any integral of the form
\[
\sint f=\frac {1} {N}\sum _{n=1} ^{N} w_n f_n~~~ (w_n>0);
\]
the axioms are trivial to verify. For this integral the state defined by
\[
\< f\> =\frac {1} {N}\ \sum _{n=1} ^{N} f_n,
\]
is a state with entropy $S$ given by $S_n=\kbar \log w_n$.
The value of the entropy
\[
\ol{S}=\frac {1} {N} \sum_{n=1}^{m} S_n =
\frac {\kbar } {N}\log \prod _{n=1} ^{N}w_n,
\]
is negative if we choose the $w_n$ such that $\prod w_n < 1 $.
\end{expl}

Thus, we need an additional condition which guarantees the validity
of the third law. 
Since the third law is also violated in classical statistical mechanics,
which is a particular case of the present setting, we need a condition
which forbids the classical interpretation of our axioms.
 
We take our inspiration from a simple information 
theoretic model of states discussed in Appendix \ref{s.complexity},
which has this property. (Indeed, the third law is a necessary 
requirement for the interpretation of the value of the entropy 
as a measure of internal complexity, as discussed there.)
There, the integral is a sum over the components, and, since functions 
were defined componentwise,
\lbeq{5-1}
\sint F(f)=\sum _{n\in {\cal N}} F(f_n).
\eeq
We say that a quantity $f$ is {\bf quantized} iff \gzit{5-1}
holds with a suitable {\bf spectrum} $\{f_n\mid n\in{\cal N}\}$ 
for all functions $F$ for which $F(f)$ is strongly integrable; 
in this case, the $f_n$ are called the {\bf levels} of $f$. 
For example, in the quantum setting 
all trace class linear operators are quantized quantities, since these 
always has a discrete spectrum. 

Quantization is the additional ingredient needed to derive the 
third law:

\begin{thm} \label{5.3.}
{\bf (Third law of thermodynamics)} \\
If the entropy $S$ is quantized then $\ol{S}\ge 0$. 
Equality holds iff the entropy has a single level only ($|{\cal N}|=1$).
\end{thm}

\bepf
We have
\lbeq{5-2}
1=\sint\rho =\sint e^{-S/\kbar }=\sum _{n\in {\cal N}}
e^{-S_n / \kbar }=\sum _{n\in {\cal N}}\rho _n,
\eeq
where all $\rho _n=e^{-Sn/\kbar }>0$, and
\lbeq{5-3}
\ol{S}=\sint S\rho =\sint Se^{-S/\kbar }
=\sum _{n\in {\cal N}}S_ne^{-S_n/\kbar }
=\sum _{n\in {\cal N}}S_n\rho _n.
\eeq
If ${\cal N}=\{n\}$ then (\ref{5-2}) implies $\rho _n=1$, $S_n=0$, and
(\ref{5-3}) gives $\ol{S}=0$. And if $|{\cal N}|>1$ then
(\ref{5-2}) gives $\rho _n<1$, hence $S_n>0$ for all $n\in {\cal N}$, 
and (\ref{5-3}) implies $\ol{S}>0$.
\epf

In quantum chemistry, energy $H$, volume $V$, and particle
numbers $N_1,\dots ,N_s$ form a quantized family of pairwise commuting
Hermitian variables. Indeed, the Hamiltonian $H$ has discrete energy 
levels if the system is confined to a finite volume, $V$ is a number,
hence has a single level only, and $N_j$ counts particles hence has 
as levels the nonnegative integers.
As a consequence, the entropy $S=T^{-1}(H+PV-
\mu \cdot N)$ is quantized, too, so that the third law of
thermodynamics is valid. The number of levels is infinite, so that
the value of the entropy is positive. 

A zero value of the entropy ({\bf absolute zero}) is therefore
an idealization which cannot be realized in practice.
But Theorem \ref{5.3.} implies in this idealized situation that
entropy and hence the joint spectrum of $(H,\ V,\ N_1,\dots ,\ N_S)$
can have a single level only. 

This is the situation discussed in ordinary
quantum mechanics (pure energy states at fixed particle numbers).
It is usually associated with the limit $T\to 0$, though at absolute 
temperature $T=0$, i.e., infinite coldness $\beta$, the thermal 
formalism fails (but low $T$ asymptotic expansions are possible).

To see the behavior close to this limit, we consider for simplicity a 
canonical ensemble with Hamiltonian $H$ (Example \ref{ex.canonical});
thus the particle number is fixed. Since $S$ is quantized, the spectrum 
of $H$ is discrete, so that there is
a finite or infinite sequence $E_0<E_1<E_2<\dots$ of distinct energy 
levels. Denoting by $P_n$ the (rank $d_n$) orthogonal projector to the 
$d_n$-dimensional eigenspace with energy $E_n$, we have the spectral 
decomposition
\[
\phi(H)= \sum_{n\ge 0} \phi(E_n) P_n
\]
for arbitrary functions $\phi$ defined on the spectrum. In particular,
\[
e^{-\beta H}= \sum e^{-\beta E_n} P_n.
\]
The partition function is 
\[
Z= \tr e^{-\beta H} = \sum e^{-\beta E_n} \tr P_n.
= \sum e^{-\beta E_n} d_n.
\]
As a consequence, 
\[
e^{-S/\kbar}= Z^{-1} e^{-\beta H} 
= \frac{\D\sum e^{-\beta E_n} P_n}{\D\sum e^{-\beta E_n} d_n}
= \frac{\D\sum e^{-\beta(E_n-E_0)}P_n}{\D\sum e^{-\beta(E_n-E_0)}d_n},
\]
hence values take the form
\lbeq{e.low}
\<f\> = \sint e^{-S/\kbar}f = \sint\Big(
 \frac{\sum e^{-\beta(E_n-E_0)}P_n}{\sum e^{-\beta(E_n-E_0)}d_n}
\Big).
\eeq
From this representation, we see that only the energy levels $E_n$
with 
\[
E_n \le E_0 +O(\kbar T)
\]
contribute to a canonical ensemble of temperature $T$.
If the temperature $T$ is small enough, so that $\kbar T \ll E_2-E_0$,
the exponentials $e^{-\beta (E_n-E_0)}$ with $n\ge 2$ can be neglected,
and we find
\lbeq{e.2levelapprox}
e^{-S/\kbar} \approx  
\frac{P_0+ e^{-\beta (E_1-E_0)} P_1}{d_0+e^{-\beta (E_1-E_0)} d_1}
= \frac{P_0}{d_0} 
+ \frac{d_0 P_1-d_1P_0}{d_0(e^{\beta (E_1-E_0)}d_0+d_1)}.
\eeq
Thus, the system behaves essentially as the two level system discussed 
in Examples \ref{ex.canonical} and \ref{ex.schottky}; the 
{\bf spectral gap} $E_1-E_0$ takes the role of $E$.
In particular, if already  $\kbar T \ll E_1-E_0$, we find that 
\[
e^{-S/\kbar} = d_0^{-1}P_0 + O(e^{-\beta (E_1-E_0)})
\approx d_0^{-1}P_0 ~~~(\mbox{if }\kbar T \ll E_1-E_0)
\]
is essentially the projector to the
subspace of minimal energy, scaled to ensure trace one.

In the {\bf nondegenerate} case, where the lowest energy eigenvalue is 
simple, there is a corresponding normalized eigenvector $\psi$, 
unique up to a phase, satisfying the {\bf Schr\"odinger equation}
\lbeq{e.schroe}
H \psi = E_0 \psi,~~|\psi|=1~~~(E_0 \mbox{ minimal}).
\eeq
In this case, the projector is $P_0=\psi\psi^*$ and has rank $d_0=1$.
Thus 
\[
e^{-S/\kbar} = \psi\psi^* + O(e^{-\beta (E_1-E_0)}).
\]
has almost rank one, and the value takes the form
\lbeq{e.quant0}
\<g\>=\tr e^{-S/\kbar} g \approx \tr \psi\psi^*g = \psi^*g\psi.
\eeq
In the terminology of quantum mechanics, $E_0$ is the {\bf ground state 
energy}, the solution $\psi$ of \gzit{e.schroe} is called the 
{\bf ground state}, and $\<g\>= \psi^*g\psi$ is the expectation of the 
observable $g$ in the ground state.

Our derivation therefore shows that -- unless the ground state is 
degenerate -- {\em a canonical ensemble at sufficiently low temperature 
is in an almost pure state described by the quantum mechanical 
ground state}. 

Thus, the third law directly leads to the conventional form of quantum 
mechanics, which can therefore be understood as the low temperature 
limit of thermodynamics. It also indicates when a quantum mechanical 
description by a pure state is appropriate, namely always when the
gap between the ground state energy and the next energy level is
significantly larger than the temperature, measured in units of the 
Boltzmann constant. (This is the typical situation in most of quantum 
chemistry and justifies the use of the Born-Oppenheimer approximation 
in the absence of level crossing; cf. {\sc Smith} \cite{Smi}, 
{\sc Yarkony} \cite{Yar}). 
Moreover, it gives the correct (mixed) form of the state
in case of ground state degeneracy, and the form of the correction
terms when the energy gap is not large enough for the ground state 
approximation to be valid.

It is remarkable that thermodynamics in this way predicts the 
Schr\"odinger equation, the relevance of the spectrum of the 
Hamiltonian, and the formula for quantum expectations in a pure state.
Indeed, after adding suitable dynamical assumptions, it is possible
to interpret all quantum mechanics from this point of view. 
The resulting {\bf thermal interpretation} of quantum mechanics will 
be discussed in {\sc Neumaier} \cite{Neu.scy}.

\section{Local, microlocal, and quantum equilibrium} \label{s.model}

As we have seen in Section \ref{s.detail}, when descriptions on 
several levels are justified empirically, they differ significantly 
only in quantities which are negligible in the more detailed models, 
or by terms which are not observable in principle. 
Thus, the global equilibrium description is adequate at some resolution 
if and only if only small nonequilibrium forces are present, and a 
more detailed local equilibrium description will (apart from variations 
of the Gibbs paradox which should be cured on the more detailed level) 
agree with the global 
equilibrium description to the accuracy within which the differences 
in the corresponding approximations to the entropy, as measured by the 
relative entropy \gzit{4-5}, are negligible. Of course, if the
relative entropy of a thermal state relative to the true Gibbs state
is large then the thermal state cannot be regarded as a faithful 
description of the true state of the system, and the thermal  
model is inadequate.

In statistical mechanics proper (where the microscopic dynamics is
given), the relevant extensive quantities are those whose 
values vary slowly enough to be macroscopically observable at a given 
spatial or temporal resolution (cf. {\sc Balian} \cite{Bal2}). 
Which ones must
be included is a difficult mathematical problem which has been solved 
only in simple situations (such as monatomic gases) where a weak 
coupling limit applies. In more general situations, the selection is 
currently based on phenomenological consideration, without any formal 
mathematical support.

In equilibrium statistical mechanics, which describes 
time-independent, {\em global} equilibrium situations, 
the relevant extensive quantities are the additive conserved quantities 
of a microscopic system and additional parameters describing order 
parameters that emerge from broken symmetries or various defects
not present in the ideal model.
{\bf Phase equilibrium} needs, in addition, copies of the extensive 
variables (e.g., partial volumes) for each phase, since the phases are 
spatially distributed, while the intensive variables are shared by all 
phases.
{\bf Chemical equilibrium} also accounts for exchange of atoms through 
a fixed list of permitted chemical reactions whose length is again 
determined by the desired resolution.

In states not corresponding to global equilibrium -- usually called 
{\bf non-equilibrium states}, a thermal description is still 
possible assuming so-called {\bf local equilibrium}. There, 
the natural extensive quantities are those whose 
values are locally additive and slowly varying in space 
and time and hence, reliably observable at the scales of interest. 
In the statistical mechanics of local equilibrium, the thermal 
variables therefore become space- and time-dependent fields 
({\sc Robertson} \cite{Rob}).
On even shorter time scales, phase space behavior becomes relevant, 
and the appropriate description is in terms of {\bf microlocal 
equilibrium} and position- and momentum-dependent phase space densities.
Finally, on the microscopic level, a linear operator description in 
terms of {\bf quantum equilibrium} is needed.

\bigskip
The present formalism is still applicable to local, microlocal, and
quantum equilibrium (though most products now become inner products in 
suitable function spaces), but the relevant quantities are now 
time-dependent and additional dynamical issues (relating states at 
different times) arise which are outside the scope of the present 
paper.

In local equilibrium, one needs a 
hydrodynamic description by Navier-Stokes equations and their 
generalizations; see, e.g., {\sc Beris \& Eswards} \cite{BerE}, 
{\sc Oettinger} \cite{Oet}, {\sc Edwards} et al. \cite {EdwOJ}. 
In the local view, one gets the interpretation of 
extensive variables as locally conserved (or at least slowly varying) 
quantities (whence additivity) and of intensive variables as 
parameter fields, which cause non-equilibrium currents when they are 
not constant, driving the system towards global equilibrium.
In microlocal equilibrium, one needs a kinetic  description by
the Boltzmann equation and its generalizations; see, e.g., 
{\sc Bornath} et al. \cite{BorKKS},
{\sc Calzetta \& Hu} \cite{CalH}, 
{\sc M\"uller \& Ruggeri} \cite{MueR}.

\bigskip
{\bf Quantum equilibrium.} 
Full microscopic dynamics must be based on quantum mechanics.
In quantum equilibrium, the dynamics is given by quantum dynamical 
semigroups. We outline the ideas involved, in order to emphasize some 
issues which are usually swept under the carpet.

Even when described at the microscopic level, thermal systems
of sizes handled in a laboratory are in contact with their environment, 
via containing walls, emitted or absorbed radiation, etc.. 
We therefore embed
the system of interest into a bigger, completely isolated system 
and assume that the quantum state of the big system is described at a 
fixed time by a normalized wave function $\psi$ in some Hilbert space 
$\widehat \Hz$. (Assuming instead a mixed state given by a density 
operator would not alter the picture significantly.)
The value of a linear operator $g$ in the big system is
\lbeq{e.quant}
\<g\> = \psi^* g \psi;
\eeq
cf. \gzit{e.quant0}.
The small system is defined by a Euclidean *-algebra $\Ez$ of linear 
operators densely defined on $\widehat \Hz$, composed of all meaningful
expressions in field operators at arguments in the region of interest, 
with integral given by the trace in the big system.
Since \gzit{e.quant}, restricted to $g\in \Ez$, satisfies the rules
(R1)--(R4) for a state, the big system induces on the system of
interest a state. By standard theorems (see, e.g.,
{\sc Thirring} \cite{Thi}), there is a unique {\bf density operator}
$\rho\in\Ez$ such 
that $\<g\> = \sint \rho g$ for all $g\in \Ez$ with finite 
value. Moreover, $\rho$ is Hermitian and positive 
semidefinite. If $0$ is not an eigenvalue of $\rho$ (which will 
usually be the case), then $\<\cdot\>$ is a Gibbs state with entropy
$S=-\kbar \log \rho$. To put quantum equilibrium into the thermal 
setting, we need to choose as extensive variables a family spanning 
the algebra $\Ez$; then each such $S$ can be written in the form 
\gzit{3-2}.

Of course, $\psi$ and hence the state $\<\cdot\>$ depend on time.
If the reduced system were goverened by a Schr\"odinger equation
then $\rho$ would evolve by means of a unitary evolution; 
in particular, $\ol S = \<S\> = -\kbar \tr \rho \log \rho$
would be time-independent. However, the system of interest does 
{\em not} inherit a Schr\"odinger dynamics from the isolated big system;
rather, the dynamics of $\rho$ is given by an 
integro-differential equation with a complicated memory term, defined 
by the projector operator formalism described in detail in 
{\sc Grabert} \cite{Gra}; for summaries, see 
{\sc Rau \& M\"uller} \cite{RauM} and {\sc Balian} \cite{Bal2}. 
In particular, one can say nothing specific about the dynamics of
$\ol S$.

In typical treatments of such reduced descriptions, one assumes that
the memory decays sufficiently fast; this so-called {\bf Markov 
assumption} can be justified in a weak coupling limit 
({\sc Davies} \cite{Dav}, {\sc Spohn} \cite{Spo}), which corresponds to
a system of interest nearly independent of the environment.
But a typical thermal system, such as a glass of water on a desk
is held in place by the container. Considered as a nearly independent 
system, the water would behave very differently, probably diffusing 
into space. Thus, it is questionable whether the Markov assumption
is satisfied; a detailed investigation of the situation would be
highly desirable. I only know of few discussions of the 
problem how containers modify the dynamics of a large quantum system;
see, e.g., {\sc  Lebowitz \& Frisch} \cite{LebF},
{\sc Blatt} \cite{Bla} and {\sc Ridderbos} \cite{Rid}.
One should expect a decoherence effect ({\sc Brune} et al. \cite{BruHD})
of the environment on the system which, for large quantum systems, 
is extremely strong ({\sc Zurek} \cite{Zur}).
A fundamental derivation should be based on quantum field theory;
the so-called exact renormalization group equations (see, e.g.,
{\sc Polonyi \& Sailer} \cite{PolS}, {\sc Berges} \cite{Ber}) have a 
thermal flavor and might be a suitable starting point. 

However, simply assuming the Markov assumption as the condition 
for regarding the system of interest to be {\bf effectively isolated} 
allows one to deduce for the resulting {\bf Markov approximation} 
a deterministic differential equation for the density operator. 
The dynamics then describes a linear quantum dynamical semigroup. 
All known linear quantum dynamical semigroup semigroups 
(cf. {\sc Davies} \cite{Dav}) on a Hilbert space correspond to a 
dynamics in the form of a {\bf Lindblad equation} 
\lbeq{e.lind}
\dot\rho= \frac{i}{\hbar} (\rho H-H^*\rho)+P^*\rho
\eeq
({\sc Lindblad} \cite{Lin}, {\sc Gorini} et al. \cite{GorKS}),
where the {\bf effective Hamiltonian} $H$ is a not necessarily 
Hermitian operator and $P^*$ is the dual of a completely positive map 
of the form 
\[
P(f) = Q^*J(f)Q \Forall f\in\Ez,
\]
with some linear operator $Q$ from $\Ez$ to a second *-algebra $\Ez'$  
and some *-algebra homomorphism $J$ from $\Ez$ to $\Ez'$.
({\sc Stinespring} \cite{Sti}, {\sc Davies} \cite[Theorem 2.1]{Dav}).
Their dynamics is inherently dissipative; for time $t\to\infty$, 
$P^*\rho$ tends to zero, which usually implies that, apart from a 
constant velocity, the limiting state is a global equilibrium state.
 
Thus, the irreversibility of the time evolution is apparent already
at the quantum level, being caused by the fact that all our observations
are done in a limited region of space. The prevalence here on earth of 
matter in approximate equilibrium could therefore possibly be 
explained by the fact that the earth and with it most of its materials 
are extremely old.

For a system reasonably isolated (in the thermodynamical sense) from 
its environment, one would expect $H$ to contain a confining effective 
potential well and $P$ to be small. It would be interesting to 
understand the conditions (if there are any) under which the 
dissipation due to $P$ can be neglected.

\bigskip 
We now consider relations within the hierarchy of the four levels. 
The quantum equilibrium entropy $S_\fns{qu}$, the microlocal 
equilibrium entropy $S_\fns{ml}$, the local equilibrium 
entropy $S_\fns{lc}$, the global equilibrium entropy $S_\fns{gl}$ 
denote the values of the entropy in a thermal description of 
the corresponding equilibrium levels. The four levels have a more and 
more restricted set of extensive quantities, and the relative entropy 
argument of Theorem \ref{t4.5} can be applied at each level. Therefore
\lbeq{e.hi1}
S_\fns{qu} \le S_\fns{ml} \le S_\fns{lc} \le S_\fns{gl}.
\eeq
In general the four entropies might have completely different
values. There are four essentially different possibilities,

 (i) $S_\fns{qu}\approx S_\fns{ml}\approx S_\fns{lc}\approx S_\fns{gl}$,

 (ii) $S_\fns{qu}\approx S_\fns{ml}\approx  S_\fns{lc} \ll S_\fns{gl}$,

 (iii) $S_\fns{qu} \approx S_\fns{ml} \ll S_\fns{lc} \le S_\fns{gl}$,

 (iv) $ S_\fns{qu} \ll S_\fns{ml} \le S_\fns{lc} \le S_\fns{gl}$,

with different physical interpretations. As we have seen in 
Section \ref{s.detail}, a thermal description is valid only if the 
entropy in this description approximates the true entropy sufficiently 
well. All other entropies, when significantly different, do not 
correspond to a correct description of the system; their disagreement 
simply means failure of the coarser description to match reality.
(Again, we disregard variations of the Gibbs paradox which should be 
cured on the fundamental level.) 
Thus which of the cases (i)--(iv) occurs decides upon which 
descriptions are valid.
(i) says that the state is in global equilibrium, and all four 
descriptions are valid.
(ii) that the state is in local, but not in global equilibrium,
and only the three remaining descriptions are valid.
(iii) says that the state is in microlocal, but not in local 
equilibrium, and in particular not in global equilibrium. 
Only the basic and the microlocal descriptions are valid.
Finally, (iv) says that the state is not even in microlocal equilibrium,
and only the quantum description is valid.

Thus (assuming that the fundamental limitations in obserability are  
correctly treated on the quantum level), the entropy is an objective 
quantity, independent of the level of accuracy with which we are able 
to describe the system, although the precise value it gets in a model 
of course depends on the accuracy of the model. The observation 
(by {\sc Grad} \cite{Grad}, {\sc Balian} \cite{Bal2}, and others) 
that entropy may depend significantly on the description level 
is explained by two facts which hold for models of any kind, not just
for entropy, namely:\\
(i) that if two models disagree in their predictions, at most one one 
of them can be correct, and\\
(ii) that if two models agree in their predictions, the more detailed
model has unobservable details.\\
Since unobservable details cannot be put to an experimental test, 
the more detailed model in case (ii) is questionable unless dictated
by fundamental considerations of formal simplicity.

\begin{appendix}

\section{Appendix: Entropy and unobservable complexity} 
\label{s.complexity}

The concept of entropy is usually introduced either historically by 
the Carnot cycle and an assumed informal form of the second law of 
thermodynamics, or, following a more recent (1957) subjectivistic 
approach ({\sc Jaynes} \cite{Jay1,Jay2}), by the information available 
to an observer and the second law in form of the principle of maximum 
entropy.

In our treatment, we avoided both, rejecting them as being
principles that require explanation themselves. 
The Carnot cycle is clearly inappropriate in a foundational 
setting, and a subjectivist approach is inappropriate for the 
description of a world that existed long before there were human 
observers.

However, to connect the traditional approach with the present setting, 
we present in this section an informal example of a simple 
stochastic model in which the entropy indeed has an information 
theoretical interpretation and then discuss what this can
teach us about a probability-free macroscopic view of the situation.

Suppose that we have a simple stationary device
which, in regular intervals, delivers a reading $n$ from a countable
set $\cal N$ of possible readings. For example, the device might
count the number of events of a certain kind in fixed periods of
time; then ${\cal N}=\{0,1,2,\dots \}$.

We suppose that, by observing the device in action for some time, we
are led to some conjecture about the (expected) relative frequencies
$p _n$ of readings $n\in \cal N$; since the device is 
stationary, these relative frequencies are independent of time. 
(If $\cal N$ is finite and not too
large, we might take averages and wait until these stabilize to a
satisfactory degree; if $\cal N$ is large or infinite, most 
$n\in \cal N$ will not have been observed, and our conjecture must 
depend on educated guesses. This introduces some subjectiveness and 
is the reason why the following material is deferred to an appendix.)

Clearly, in order to have a consistent interpretation of the $p _n$ 
as relative frequencies, we need to assume that {\em each} reading is 
possible:
\lbeq{1-1}
p_n > 0 \mbox { for all }n\in \cal N,
\eeq
and {\em some} reading occurs with certainty:
\lbeq{1-2}
\sum _{n\in \cal N}p _n=1.
\eeq
(For reasons of economy, we shall not allow $p_n=0$ in \gzit{1-1},
which would correspond to readings that are either impossible,
or occur too rarely to have a scientific meaning. Clearly, this is no
loss of generality.)
Knowing relative frequencies only means that (when ${\cal N}>1$) we only
have incomplete information about future readings of the device.
We want to calculate the information deficit by counting the expected
number of questions needed to identify a particular reading unknown
to us, but known to someone else who may answer our questions with
yes or no.

Consider an arbitrary strategy for asking questions, and denote by
$s_n$ the number of questions needed to determine the reading $n$.
With $q$ questions we can distinguish up to $2^q$ different cases; but
since reading $n$ is already determined after $s_n$ questions,
reading $n$ is obtained in $2^{q-s_n}$ of the $2^q$ cases (when $s_n\le
q$). Thus
\[
\sum _{s_n\le q}2^{q-s_n}\le 2^q.
\]
If we divide by $2^q$ and then make $q$ arbitrarily large we find that
\lbeq{1-3}
\sum _{n\in \cal N}2^{-s_n}\le 1.
\eeq
It is not difficult to construct strategies realizing the $s_n$
whenever (\ref{1-3}) holds. 

Since we do not know in advance the reading,
we cannot determine the precise number of questions needed in a
particular unknown case. However, knowledge of the relative frequencies
allows us to compute the average number of questions needed, namely
\lbeq{1-4}
\ol{s}=\sum _{n\in \cal N}p_n s_n.
\eeq
To simplify notation, we introduce the abbreviation
\lbeq{1-5}
\sint  f:=\sum _{n\in \cal N}f_n
\eeq
for every quantity $f$ indexed by the elements from $\cal N$, and we 
use the convention that inequalities, operations and functions of such 
quantities are understood componentwise. Then we can rewrite 
\gzit{1-1}--\gzit{1-4} as
\lbeq{1-6}
p >0,\quad \sint p=1\ ,
\eeq
\lbeq{1-7}
\ol{s}=\sint p s,\quad \sint 2^{-s}\le 1\ ,
\eeq
and
\lbeq{1-8}
\ol{f}=\<f\>:=\sint p f.
\eeq
is the average of an arbitrary quantity $f$ indexed by $\cal N$.

\bigskip
We now ask for a strategy which makes the number $\ol{s}$
as small as possible. However, we idealize the situation a little by 
allowing the $s_n$
to be arbitrary nonnegative real numbers instead of integers only.
This is justified when the size of $\cal N$ is large or infinite
since then most $s_n$ will
be large numbers which can be approximated by integers with a tiny
relative error.

\begin{thm} \label{1.1.}

The {\bf entropy} $S$, defined by
\lbeq{1-9}
S:=-\kbar \log p, ~~~\mbox{where}~~\kbar=\frac {1} {\log 2},
\eeq
satisfies $\ol{S}\le \ol{s}$, with equality if and only if $s=S$.
\end{thm}

\bepf
(\ref{1-9}) implies $\log p =-S\log 2$, hence $p =2^{-S}$.
Therefore
$$
2^{-s}=p 2^{S-s}=p e^{\log 2(S-s)}\ge p
(1+\log 2(S-s)),
$$
with equality iff $S=s$. Thus
$$
p (S-s)\le \frac {1} {\log 2}(2^{-s}-p )=\kbar (2^{-s}-p )
$$
and
\[ 
\begin{array}{lll} 
\ol{S}-\ol{s}&=\sint p (S-s)\le \sint 
\kbar (2^{-s}-p ) \\
 &=\kbar~\sint 2^{-s}-\kbar ~ \sint p 
\le \kbar -\kbar =0.
\end{array} 
\]
Hence $\ol{s}\ge \ol{S}$, and equality holds iff $s=S$.
\epf

Since \gzit{1-9} implies the relation $p =e^{-S/\kbar }$, we have
$\<f\>=\sint p f= \sint e^{-S/\kbar }f$. Thus, the expectation mapping 
is a Gibbs state with entropy $S$, explaining the name.
Note that $s=S$ defines an admissible strategy since
$$
\sum _{n\in \cal N}2^{-S_n}=\sint 2^{-S}=\sint p =1, 
$$
hence $2^{-S_n}\le 1$, $S_n\ge 0$ for all $n\in \cal N$. 
Thus, the entropy $S$ is {\bf the unique optimal decision strategy.}
The {\bf expected entropy}, i.e., the mean number
\lbeq{1-10}
\ol{S}=\<S\>=\sint pS =-\kbar~ \sint p \log p 
\eeq
of questions needed in an optimal decision strategy, is nonnegative,
\lbeq{1-11}
\ol{S}\ge 0.
\eeq
It measures the {\bf information deficit} of the device with respect 
to our conjecture about relative frequencies. (Traditionally, this is 
simply called the entropy, while we reserve this word for the random 
variable \gzit{1-9}.
Also commonly used is the name {\em information} for $\ol{S}$, 
which invites linguistic paradoxes since ordinary language associates 
with information a connotation of 
relevance or quality which is absent here. 
An important book on information by {\sc Brillouin} \cite{Bri} 
emphasizes this very carefully, by distinguishing absolute information 
from its human value or meaning. {\sc Katz} \cite{Kat} uses the 
phrase 'missing information'.) 

The information deficit says nothing at all about the quality
of the information contained in the summary $p $ of our past 
observations.
An inappropriate $p $ can have arbitrarily small information
deficit and still give a false account of reality. E.g., if for some 
small $\eps>0$,
\lbeq{1-10a}
 p _n=\eps^{n-1}(1-\eps) \quad \mbox{for}~ n=1,2,\dots,
\eeq
expressing that the reading is expected to be
nearly always 1 ($p_1=1-\eps$) and hardly ever large, then
$$
 \ol{S}=\kbar \Big(\log(1-\eps )+\frac{\eps}{1-\eps}\log\eps\Big) 
\to 0 ~\mbox{as} ~\eps \to 0. 
$$
Thus the information deficit
can be made very small by the choice \gzit{1-10a} with small $\eps$, 
independent of whether this choice corresponds to the known facts. The
real information value of $p $ depends instead on the care 
with which the past observations were interpreted, which is a matter of
data analysis and not of our model of the device. If this is done badly,
our expectations will simply not be matched by reality.
This shows that the entropy has nothing to do with ``our knowledge of 
the system'' -- a subjective, ill-defined notion -- but reflects 
objective properties of the stochastic process.

\bigskip
{\bf Relations to thermodynamics.}
Now suppose that the above setting happens at a very fast, unobservable 
time scale, so that we can actually observe only short time 
averages (\ref{1-8}) of quantities of interest. Then $\ol f=\<f\>$ 
simply has the interpretation of the time-independent observed value 
of the quantity $f$. The information deficit
simply becomes the observed value of the entropy $S$. 
Since the information deficit counts the number of optimal 
decisions needed to completely specify a (microscopic)
situation of which we know only (macroscopic) observed values,
the observed value of the entropy quantifies the {\bf intrinsic} 
(microscopic) {\bf complexity} present in the system.

However, the unobservable high frequency fluctuations of the device 
do not completely disappear from the picture. 
They show in the fact that generally $\ol{g^2}\ne\ol g^2$, 
leading to a nonzero limit resolution \gzit{e.res} of Hermitian 
quantities.
This is precisely the situation characteristic of the traditional 
treatment of thermodynamics within classical equilibrium statistical 
mechanics, if we assume {\em ergodicity}, i.e., that ensemble averages 
equal time averages. (This is a problematic assumption; see, e.g., 
the discussion in {\sc Sklar} \cite{Skl}.) There, all observed values 
are time-independent, described by equilibrium thermal 
variables.
But the underlying high-frequency motions of the atoms making up 
a macroscopic substance are revealed by nonzero limit resolutions.

Note that even a deterministic but chaotic high frequency dynamics, 
viewed at longer time scales,
looks stochastic, and exactly the same remarks about the unobservable
complexity and the observable consequences of fluctuations apply. 
Even if fluctuations are observable directly, these observations are 
intrinsically limited by the necessary crudity of any actual 
measurement protocol. For the best possible measurements (and only for
these), the resolution of $f$ in the experiment is given by the limit
resolution $\res(f)$, the size of the unavoidable fluctuations.

Due to the quantum structure of high frequency phenomena (on an atomic 
or subatomic scale), it may, however, seem problematic to interpret the 
thermodynamic limit resolutions in terms of a simple short time average 
of some underlying microscopic reality. Fortunately, as we have seen, 
such an interpretation is not necessary.

\section{Appendix: The maximum entropy principle} 
 \label{s.maxent}

Motivated by a subjective, information theoretic approach to 
probability, {\sc Jaynes} (\cite{Jay1} for the classical case 
and \cite{Jay2} for the quantum case), used the maximum 
entropy principle to derive the thermal formalism. 
This approach has gained considerable acceptance in the physics
community.
But why should nature be concerned about the the amount of 
information an observer has? According to whose knowledge should
it behave? Since thermodynamics is completely observer independent,
its foundations should have this feature, too.

The present approach avoids this subjective touch and shows that a 
fully objective foundation is possible. The maximum entropy principle
becomes a theorem, valid (only) under precisely specified conditions.
To denote the extensive variables, we
 use the barless notation of Section \ref{s.phen}.
\begin{thm} \label{4.1.} ({\bf Entropy form of the second law})\\
In an arbitrary state of a standard thermal system 
\[
S \le S(H,V,N)
:=\min\,\{T^{-1}(H+PV-\mu\cdot N)\mid \Delta(T,P,\mu)=0\},
\]
with equality iff the state is an equilibrium state.
The remaining thermal variables are then given by
\lbeq{e.ent1x}
T^{-1}=\frac{\partial S}{\partial H}(H,V,N),~~~
T^{-1}P=\frac{\partial S}{\partial V}(H,V,N),~~~
T^{-1}\mu=-\frac{\partial S}{\partial N}(H,V,N),
\eeq
\lbeq{e.ent2x}
U=H=TS(T,V,N)-PV+\mu\cdot N.
\eeq
\end{thm}

\bepf
This is proved in the same way as Theorem \ref{t.extstd};
thus we give no details.
\epf

The {\em only} situation in which the value of the entropy 
{\em must} increase to reach equilibrium is when $H$, $V$ and $N$ are 
kept constant. 
Under different constraints, the entropy is no longer maximal. 
For example, if one pours milk into one's coffee, stirring mixes coffee
and milk, thus increasing complexity. Macroscopic order is
restored after some time when this increased complexity has become
macroscopically inaccessible -- since $T,P$ and $N$ are constant in a 
state of minimal Gibbs energy, and not in a state of maximal entropy!
More formally, the first law shows that, for standard systems at fixed 
value of the particle number, the value of the entropy decreases 
when $\ol{H}$ or $V$ 
(or both) decrease reversibly; this shows that the value of the entropy 
may well decrease if accompanied by a corresponding decrease of 
$\ol{H}$ or $V$. The same holds out of equilibrium (though our 
argument no longer applies); for example, the reaction 
2 H${}_2$ $+$ O${}_2$ $\to$ 2 H${}_2$O (if catalyzed) 
may happen spontaneously at constant $T=25\,^\circ$C and $P=1$~atm, 
though it decreases the entropy. 

The conditions of constant $H$, $V$ and $N$, needed to argue 
that entropy must increase, are not easily realized in nature. 
From a fundamental point of view, they are even always violated:
The only truly isolated system is the universe as a whole; 
but the universe expands, i.e., $V$ changes. 
Thus, the assumption of constant $H$, $V$ and $N$ is unrealistic, and 
while the second law in the form of a maximum entropy principle may be 
of theoretical and historical importance, it is not the extremal 
principle ruling nature. 

The irreversible nature of physical processes is instead manifest, 
irrespective of the entropy balance, as energy dissipation which, 
in a microscopic interpretation, indicates 
the loss of energy to the unmodelled microscopic degrees of freedom.
Macroscopically, the global equilibrium states are therefore states 
of least free energy (the correct choice of which depends on the 
boundary condition), with the least possible freedom for change. 
(This macroscopic immutability is another intuitive explanation for the 
maximal macroscopic order in global equilibrium states.)

\bigskip
The maximum entropy principle is also heralded as an rational, 
unprejudiced way of accounting for available information in 
incompletely known statistical models. 
However, it accounts only for information about 
exact expectation values, and the model produced by the maximum entropy 
principle describes the true situation correctly only if the 
expectations of the sufficient statistics of the true model are 
available exactly; see, e.g., {\sc Barndorff-Nielsen} \cite{BarN}.
Which statistics can be considered sufficient depends on the true 
situation and is difficult to assess in advance. The maximum entropy
principle amounts in this situation simply to the (often unfounded)
assumption that the sufficient statistics are among the quantities of
which one happens to know the expectation values. Moreover, questions 
about the uninformative prior which must be assumed to describe the
state of complete ignorance affect the results of the maximum entropy 
principle, making the application of the principle ambiguous.

\begin{expls} 
(i) If we have information in the form of a large but finite sample of 
$N$ realizations $g(\Omega_k)$ ($k=1,\dots,n$) of a random variable $g$,
we can obtain from it approximate information about all moments 
$\<g^n\>\approx \sum g(\Omega_k)^n/N$ ($n=0,1,2,\dots$).
The maximum entropy principle would infer that the distribution of 
$g$ is discrete, namely that of the sample.

(ii) If we take as uninformative prior for a real-valued random 
variable $g$ the Lebesgue measure and only know that the mean of $g$
is 1, the maximum entropy principle does not produce a sensible 
probability distribution. If we add the knowledge of the second 
moment $\<g^2\>=2$, we get a Gaussian distribution with mean 1 and 
standard deviation 1. If we then get to know that the random variable 
is in fact nonnegative and integer-valued, this cannot be accounted 
for by the principle, and the probability of obtaining a negative 
value remains large. 

(iii) But if we take as prior the discrete measure on nonnegative 
integers, the 'noninformative' prior has become much more informative,
the knowledge of the mean produces the Poisson distribution, and 
the knowledge of the second moment may modify this further.

(iv) If we know that a random variable $g$ is nonnegative and has 
$\<g^2\>=1$; the  Lebesgue measure on $\Rz_+$ as noninformative prior
gives for $g$ a distribution with density $\sqrt{2/\pi}e^{-g^2/2}$.
But we can consider instead our knowledge about $h=g^2$,
which is nonnegative and has $\<h\>=1$; the same noninformative
prior now gives for $h$ a distribution with density $e^{-h}$.
The distribution of $g=\sqrt{h}$ resulting from this has density
$2ge^{-g^2/2}$. Thus the result depends on whether we regard $g$
or $h$ as the relevant variable.
\end{expls} 

These examples clearly demonstrate that the maximum entropy principle 
is an unreliable tool. The prior, far from being uninformative,
reflects the prejudice assumed without any knowledge, and the
choice of expectations to use reflects prior experience about which
expectations are likely to be relevant. Thus the application of the
maximum entropy principle becomes reliable only if one knows the 
desired for mof the result beforehand (which was indeed the case with 
Jaynes' arguments, many years after Gibbs). Thus, the principle is
not a suitable basis for the foundations of thermodynamics.

\section{Appendix: Some mathematical lemmas} \label{s.technical}

In this section we prove some mathematical results needed in the main 
text. The proofs are rigorous for the case 
$\Ez\subseteq \Cz^{n\times n}$ only; this covers the $N$-level quantum 
system, but also quantum field theory in the finite lattice 
approximation, and implies the results whenever $\Ez$ is 
finite-dimensional. 
Similar arguments work in more general situations if we use spectral 
resolutions; cf., e.g., {\sc Thirring} \cite{Thi} (who works in 
$C^*$-algebras and von Neumann algebras). I'd appreciate to be 
informed about possible proofs in general that only use the properties 
of Euclidean *-algebras (and perhaps further, elementary assumptions).

\bigskip
\begin{prop} \label{app2a.} 
For arbitrary quantities $f$, $g$,
\[
e^{\alpha f}e^{\beta f}=e^{(\alpha+\beta)f}~~(\alpha,\beta\in\Rz),
\]
\[
(e^f)^*=e^{f^*},
\]
\[
e^f g = g e^f ~~~\mbox{if $f$ and $g$ commute},
\]
\[
f^*=f \implies \log e^f=f,
\]
\[
f\ge 0 \implies \sqrt{f}\ge 0,~~(\sqrt{f})^2 =f,
\]
For any quantity $f=f(s)$ depending continuously on $s\in[a,b]$,
\[
\int_a^b ds \sint f(s) = \sint \Big(\int_a^b ds f(s)\Big),
\]
and for any quantity $f=f(\lambda)$ depending continuously 
differentiably on a parameter vector $\lambda$, 
\[
\frac{d}{d\lambda} \sint f = \sint df/d\lambda.
\]
\end{prop}

\bepf 
In finite dimensions, the first four assertions are standard 
matrix calculus, and the remaining two statements hold since $\sint f$ 
must be a finite linear combination of the components of $f$.
\epf

\begin{prop} \label{app1.}
Let $f,g$ be quantities depending continuously differentiably on a
parameter vector $\lambda $, and suppose that
\[
[f(\lambda ),g(\lambda )]=0\mbox { for all }\lambda.
\]
Thus, for any continuously differentiable function $F$ of two
variables,
\lbeq{app1}
\frac {d} {d\lambda }\sint F(f,g)
=\sint\partial _1F(f,g)\frac {df} {d\lambda }
+\sint\partial _2F(f,g)\frac {dg } {d\lambda }\ .
\eeq
\end{prop}

\bepf
We prove the special case $F(x,y)=x^my^n$, where (\ref{app1}) reduces
to
\lbeq{app2}
\frac {d} {d\lambda }\sint f^mg^n
=\sint mf^{m-1}g^n\frac {df} {d\lambda }
+\sint nf^mg^{n-1}\frac {dg} {d\lambda}.
\eeq
The general case then follows for polynomials $F(x,y)$ by taking
suitable linear combinations, and for arbitrary $F$ by a limiting
procedure. To prove (\ref{app2}), we note that, more generally,
\[ 
\begin{array}{lll}
\D\frac {d} {d\lambda }\sint f_1\dots f_{m+n}
&=\sint\frac {d} {d\lambda }(f_1\dots f_{m+n})\\
&\D=\sint\sum _{j=1} ^{m+n}f_1\dots f_{j-1}\frac {df_j} {d\lambda }
f_{j+1}\dots f_{m+n} \\
&\D=\sum _{j=1} ^{m+n}\sint f_1\dots f_{j-1}\frac {df_j} {d\lambda }
f_{j+1}\dots f_{m+n} \\
&\D=\sum _{j=1} ^{m+n}\sint f_{j+1}\dots f_{m+n}f_1\dots f_{j-1}
\frac {df_j} {d\lambda }\ , 
\end{array}
\]
using the cyclic commutativity (EA2) of the integral.
If we specialize to $f_j=f$ if $j\le m$, $f_j=g$ if $j>m$, and note
that $f$ and $g$ commute, we arrive at (\ref{app2}).
\epf

Of course, the proposition generalizes to families of more than two
commuting quantities; but more important is the special case $g=f$:

\begin{cor} \label{app2.}
For any quantity $f$ depending continuously differentiably on a
parameter vector $\lambda $, and any continuously differentiable 
function $F$ of a single variable,
\lbeq{app3}
\frac {d} {d\lambda }\sint F(f)=\sint F'(f)\frac {df} {d\lambda }.
\eeq
\end{cor}

A real-valued function $\phi$ is convex in a convex set
$X\subseteq \Rz^n$ if $\phi$ is defined on $X$ and, for all $x,y\in X$,
\[
\phi(tx+(1-t)y) \le t\phi(x)+(1-t)\phi(y) \for 0\le t\le 1.
\]
Clearly, $\phi$ is convex iff for all $x,y\in X$, the function 
$\mu:[0,1]\to \Rz$ defined by
\[
\mu(t):=\phi(x+t(y-x))
\]
is convex. It is well-known that, for twice continuously 
differentiable $\phi$, this is the case iff the second derivative 
$\mu''(t)$ is nonnegative for $0\le t\le 1$. 
Note that by a theorem of Aleksandrov (see {\sc Aleksandrov} \cite{Ale},
{\sc Franklin} \cite{Fra}, {\sc Rockafellar} \cite{Roc}),
convex functions are almost everywhere twice continuously 
differentiable, in the sense that, for every $x\in X$, there exist a 
gradient vector $\frac{\partial}{\partial x}\phi(x)\in \Rz^n$ and a 
symmetric, positive definite Hessian matrix 
$\frac{\partial^2}{\partial x^2}\phi(x)\in \Rz^{n\times n}$ such that, 
for arbitrary $h \in \Rz^n$,
\[
\phi(x+h)=\phi(x)+h^T\frac{\partial}{\partial x}\phi(x)
+\half h^T\frac{\partial^2}{\partial x^2}\phi(x)h + o(\|h\|^2). 
\]
A function $\phi$ is concave iff $-\phi$ is convex. Thus, for twice 
continuously differentiable $\phi$, $\phi$ is concave iff 
$\mu''(t)\le 0$ for $0\le t\le 1$. 

\begin{prop}\label{e.convex}
If $\phi$ is convex then the function $\psi$ defined by
\[
\psi(s,x):=s\phi(x/s)
\]
is convex for $s>0$ and concave for $s<0$.
\end{prop}

\bepf
It suffices to show that $\mu(t):=\psi(s+tk,x+th)$ is convex (concave)
for all $s,x,h,k$ such that $s+tk>0$ (resp. $<0$).
Let $z(t):=(x+th)/(s+tk)$ and $c:=sh-kx$. Then
\[
z'(t)=\frac{c}{(s+tk)^2},~~~\mu(t)=(s+tk)\phi(z(t)),
\]
hence
\[
\mu'(t)=k\phi(z(t))+\phi'(z(t))\frac{c}{s+tk},
\]
\[
\mu''(t)=k\phi'(z(t))\frac{c}{(s+tk)^2}
+\frac{c^T}{(s+tk)^2}\phi''(z(t))\frac{c}{s+tk}
+\phi'(z(t))\frac{-ck}{(s+tk)^2}
=\frac{c^T\phi''(z(t))c}{(s+tk)^3},
\]
which has the required sign.
\epf

\end{appendix}

\bigskip

\end{document}